# CLARIFYING DEAD QUASARS AND DOWNSIZING THROUGH THE EVOLUTION OF STAR-FORMING GALAXIES AND ACTIVE GALACTIC NUCLEI

SHINGO SUMIE

## ABSTRACT

From the catalogs of visible emission lines of galaxies measured by the Subaru Fiber Multi-Object Spectrograph (Subaru-FMOS) and the KECK Multi-Object Spectrograph For InfraRed Exploration (KECK-MOSFIRE), galaxies were classified into Star-Forming Galaxies (SFGs) and Active Galactic Nuclei (AGNs) based on the Baldwin–Phillips–Terlevich (BPT) diagram. In the redshift (z) range $0.8 \leq z \leq 2.6$, z dependence of their number densities was examined, and it was found that SFGs dominate at lower redshifts, whereas AGNs are distributed at higher redshifts. Infrared analyses using the James Webb Space Telescope (JWST) NIRSpec and MIRI for galaxies up to $z \simeq 6.6$, show similar trends, and the number densities of SFGs and AGNs were found to cross at $z \simeq 1–3$. Combined with the fact that the luminosity of galaxies decreases monotonically with the age of the universe, these phenomena can be interpreted by an evolutionary scenario in which SFGs merged to form quasars in the early universe, whose activity weakened and they changed into SFGs later, SFGs merged again to form Seyferts in the nearby universe. Within this framework, the long-standing issues of "dead quasars" and "downsizing" can be naturally explained.

## 1. INTRODUCTION

SuperMassive Black Holes (SMBHs) with masses ranging from one million to over one billion times the mass of the Sun ($10^6$ M$_\odot$ to over $10^9$ M$_\odot$) are present at the centers of almost all galaxies, excluding dwarf galaxies like the Magellanic Clouds. The radiation of Active Galactic Nuclei (AGNs), such as quasars (Quasi-Stellar Objects: QSOs), Seyfert galaxies, and LINERs (Low-Ionization Nuclear Emission-line Regions), is attributed to the accretion of surrounding gas onto the accretion disk around the SMBH. This process releases gravitational energy, converting it into electromagnetic radiation and jet energy with an efficiency of 10%, which is significantly higher than the 0.7% efficiency of thermonuclear fusion (Lynden-Bell 1969). However, many aspects, such as the differences in activity between Star-Forming Galaxies (SFGs) in the nearby universe and AGNs, the formation processes of galaxies and SMBHs in the early universe, and the co-evolution of galaxies and SMBHs (Magorrian et al. 1998) remain unresolved (Taniguchi 2014).

Two major unresolved issues in galaxy evolution have been highlighted (Taniguchi et al. 2012):

The Dead Quasar Problem: Quasars, the brightest type of AGN, were abundant in the distant universe 2–3 billion years after the Big Bang (redshift z=2–3) but are almost absent in the nearby universe ($z < 0.1$) after 10 billion years (Fan et al. 2001). Since SMBHs with masses exceeding $10^9$ M$_\odot$ do not disappear within 10 billion years (Hawking 1975), the absence of quasars in the nearby universe is puzzling. The typical luminosity of quasars ranges from $10^{45}$ erg/s to $10^{49}$ erg/s, corresponding to absolute magnitudes of -23.7 to -33.7. Observations of the number density of galaxies and their brightness variations with redshift (Ueda et al. 2003; Ikeda et al. 2012; Hasinger et al. 2005) reveal that quasars peaked at z=2–3 and have since declined, becoming nearly absent in the nearby universe (Figure 1).

The Downsizing Phenomenon: Contrary to the natural expectation that smaller galaxies merge to form larger ones during galaxy evolution, observations indicate that the number density of brighter (larger) galaxies peaked earlier in the universe's history, while fainter (smaller) galaxies became more prevalent in the present (Kriek et al. 2007; Ueda et al. 2003; Hasinger et al. 2005; Ikeda et al. 2011; Ikeda et al. 2012). This suggests that "larger galaxies formed earlier," introducing a significant contradiction in our understanding of galaxy evolution.

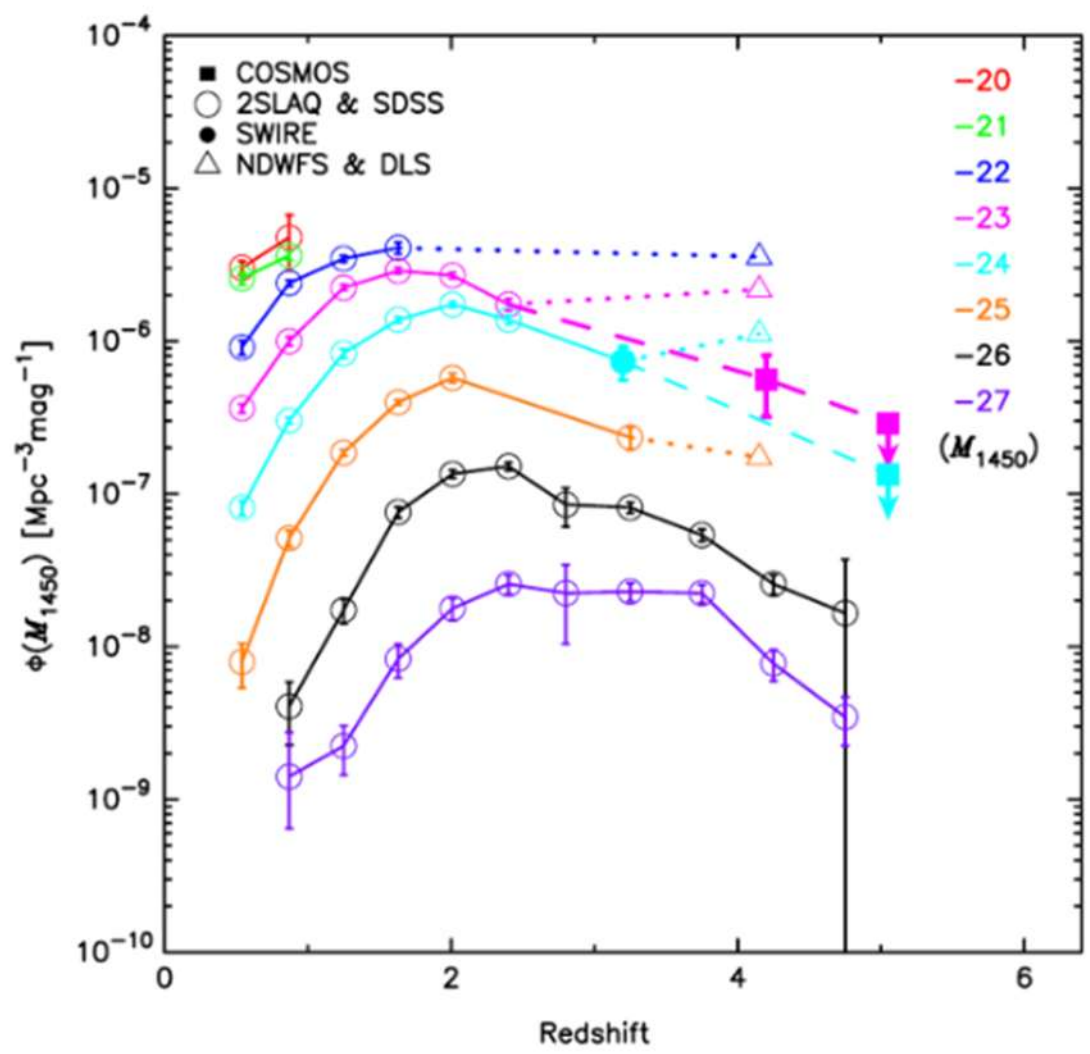

**Figure 1.** Dependence of number density of galaxies on redshift z for each absolute magnitude (Ikeda et al. 2012 Figure 9)

Based on the observational data, the macroscopic evolution of galaxies is explained through Pure Luminosity Evolution (PLE) and Pure Density Evolution (PDE). The phenomenon of decreasing number density of bright AGNs as the redshift decreases is interpreted as "the luminosity of AGNs decreases (PLE)" and "the number of AGNs decreases (PDE)". In reality, both PLE and PDE are considered to occur simultaneously. On the other hand, the mechanisms driving galaxy evolution are thought to involve the following factors: (Taniguchi 2004)

**Factor (A):** "SFGs merge and transform into AGNs."

Major mergers: Collisions of massive spiral galaxies causes intense ultra-starbursts as interstellar gas collides, and the resulting massive stars heat the gas, forming ultra-luminous infrared galaxies (ULIRGs) that emit strong infrared radiation. The gas dispersion from supernova explosions gives rise to quasar-class high-luminosity AGNs.

Minor mergers: Collisions between spiral galaxies and satellite galaxies cause the satellite galaxy's core with gas to fall into the center of spiral galaxy, inducing starbursts in high-density gas regions and forming medium-luminosity AGNs such as Seyfert galaxies.

**Factor (B):** "AGN activity decreases."

Gas depletion: AGNs are illuminated by the accretion of gas onto the SMBH (accretion disk) at a rate of about 2 $M_{\odot}$/year. Since the central gas amounts to tens of thousands to hundreds of millions of $M_{\odot}$, the gas depletes within tens of thousands to hundreds of millions of years.

ADAF (Advection-Dominated Accretion Flow): When ADAF forms, the heat generated by the release of gravitational energy is transported to the SMBH without being lost through radiation, causing the radiative luminosity to decrease (Ichimaru 1977),.

This study aims to investigate which of the mechanisms of factors (A) and (B) is dominant by analyzing the distribution trends of SFGs and AGNs across various redshifts in the universe. The goal is to interpret the problems for "Dead Quasar " and "Downsizing", which are the unresolved issues in the research history of galaxy evolution. In this paper, to avoid cumbersome terminology, SFG is defined as "star-forming galaxies or inactive galaxies," and AGN as "active galactic nuclei or host galaxies with active galactic nuclei at their center." Through this study, the density parameters were set as $\Omega_m$ (baryons + dark matter) = 0.3, $\Omega_\Lambda$ (dark energy) = 0.7, and the Hubble constant $H_0$ = 70 $kms^{-1}Mpc^{-1}$ (Toshifumi Futamase, 2002).

## 2. METHODS

### *2.1 Characteristics of Radiation from SFGs and AGNs*

From the perspective of activity, galaxies can be characterized as follows: AGNs are galaxies that exhibit activity by emitting strong electromagnetic radiation from their hot accretion disks, while SFGs are galaxies that do not exhibit such activity. The characteristics of radiation from SFGs and AGNs are summarized in Table 1, categorized by continuum light, hydrogen recombination lines, and highly ionized emission lines.

In AGNs, a broad-line region (BLR) with a width of approximately 0.1 pc is formed around the accretion disk, consisting of high-speed (high-temperature) and high-density gas plasma. Due to the Doppler effect caused by the high-speed gas movement, broad lines (BL) with a full width at half maximum (FWHM) of 5,000 km/s (500 km/s to 10,000 km/s) are emitted. Narrow-line regions (NLRs) are formed at distances of 100 pc to 1 kpc from the accretion

**Table 1.** Characteristics of continuous light, hydrogen atom recombination lines, and higher-order ionization lines emitted by SFG and AGN (Type 1, Type 2).

| Electromagnetic Radiation | | SFG | AGN | | Notes |
|---|---|---|---|---|---|
| | | | Type 1 | Type 2 | |
| Continuous Light | | Black Body Radiation | Synchrotron Radiation | | AGN radiation follows a power law. |
| Broad Line | Hydrogen Atom Recomb. Line | None (slow) | Hβ, etc. | None (torus shield) | Type 1 BLRs emit broad emission lines such as Hβ due to high speeds (>5,000 km/s) and high temperature. |
| | Higher-Order Ionization Line | None (low temp.) | $C_{IV}$, $C_{III}$], etc. | None (torus shield) | Recombination lines and semi-forbidden lines are emitted. Due to the large Doppler shifts, the both spectrums are mixed. |
| Narrow Line | Hydrogen Atom Recomb. Line | Lyα, Hα, Hβ, Hγ, etc. | Lyα, Hα, Hβ, Hγ, etc. | Lyα, Hα, Hβ, Hγ, etc. | Lyman and Balmer series due to hydrogen atom gas extending from the center to about 1 kpc. |
| | Higher-Order Ionization Line | None | $C_{IV}$, $Mg_{II}$, [$O_{III}$], [$N_{II}$], [$S_{II}$], [$O_{I}$], etc. | $C_{IV}$, $Mg_{II}$, [$O_{III}$], [$N_{II}$], [$S_{II}$], [$O_{I}$], etc. | Due to low density (2,000 $cm^{-3}$) of NLR for AGN, acceptable lines, semi-forbidden lines], and [forbidden lines] are radiated. |

BLR : Broad Line Region (Recomb. lines and semi-forbidden lines] with broad spectrum are emitted)
NLR : Narrow Line Region (Recomb. lines and semi-forbidden lines] with narrow spectrum are emitted)
Details of high-order ionization : $C_{IV}$1549, $C_{III}$]1909, $Si_{IV}$+$O_{IV}$]（$Si_{IV}$1394, 1403, $O_{IV}$]1402）, $Mg_{II}$2798, [$O_{III}$]5008, [$N_{II}$]6585, [$S_{II}$]6718, [$O_{I}$]6302 disk, consisting of low-speed (low-temperature) and low-density gas plasma. These regions emit narrow lines (NL) with an FWHM of 400 km/s (200 km/s to 900 km/s). Type 1 AGNs are observed as a face-on galaxy, allowing the BLR to be visible, and both BL and NL emission lines appear. Type 2 AGNs are observed from an edge-on direction, where the BLR is obscured by a dust torus, and only NL emission lines appear. Most quasars are Type 1, with Type 2 accounting for only about 0.1%, whereas Seyfert galaxies have a reverse ratio, with 30% being Type 1 and 70% being Type 2 (Taniguchi 2004). This is thought to be because quasars have strong radiation that destroys the torus, or because high-luminosity AGNs have geometrically thinner toruses, making the BL from the BLR visible in quasars (Peterson 2010). On the other hand, SFGs do not have active accretion disks, so the gas is low-speed and low-temperature, and the emitted emission lines are primarily narrow hydrogen lines.

### *2.2 Classification of SFGs and AGNs using BPT Diagrams with Redshift Dependence*

Methods to classify SFGs and AGNs from these emission line characteristics include using the difference in the width of Lyα emission lines (SFG: FWHM ≤ 1,000 km/s, AGN: FWHM > 1,000 km/s) (Zhang et al. 2021), detecting Lyman breaks in high-redshift quasars in two-color diagrams (Ikeda et al. 2011, Ikeda et al. 2012), and using BPT diagrams (Baldwin et al. 1981) based on visible light emission line diagnostics. In this study, the BPT diagram was chosen because it can separate SFGs and AGNs and further classify AGNs into Seyfert galaxies and LINERs.

The BPT diagram utilizes the fact that AGNs have higher fluxes (energy per unit area per unit time) of highly ionized emission lines compared to SFGs. It plots the flux ratios of forbidden high-ionization emission lines and hydrogen Balmer emission lines in a two-dimensional diagram, with regions separated by boundary lines to classify SFGs and AGNs. A commonly used diagram takes the flux ratio of the highly ionized forbidden line $[O_{III}]5008$ to the Hβ4863 line on the vertical axis and the flux ratio of the low-ionization forbidden line $[N_{II}]6585$ to the Hα6565 line on the horizontal axis.

To understand the evolution of galaxies, the redshift dependences of the number density of SFGs and AGNs were investigated across a wide range of redshifts in the universe, from $0 < z < 6.5$ (from the present to 1 billion years after the birth of the universe).

## 3. SAMPLES

When the redshift exceeds 0.5, the Balmer emission lines shift to the near-infrared region, making it necessary to perform spectroscopic observations in the near-infrared region to create BPT diagrams for the Hα–Hβ region. Due to the high redshift, the apparent brightness of galaxies becomes dimmer, and the sensitivity of detectors in the near-infrared is lower than in visible light, making precise observations challenging.

Ground-based observations require large-aperture telescopes in the range of 8m–10m that can collect a significant amount of light. Additionally, wide-field observations using multi-object spectroscopy with optical fibers or similar technologies are desirable to simultaneously observe as many celestial objects as possible. Telescopes/instruments that meet these conditions include the Subaru Telescope's Fiber Multi-Object Spectrograph (Subaru-FMOS) and the Keck Telescope's Multi-Object Spectrograph For InfraRed Exploration (KECK-MOSFIRE), therefore, data from these two databases were analyzed.

The James Webb Space Telescope (JWST), launched at the end of 2021, is primarily exploring the high-redshift universe. In this study, two arXiv papers based on JWST observation data were analyzed.

## 4. RESULTS

### *4.1 Boundary Settings in BPT Diagram and Analysis of Kartaltepe et al. 2015 (Subaru-FMOS)*

The methods for setting the boundary between SFGs and AGNs in the BPT diagram include the fixed boundary method (Brinchmann et al. 2004) and the variable boundary method (Kewley et al. 2013a). However, since there is little difference between the two methods for understanding the redshift dependence of the number density of SFGs and AGNs, the fixed boundary method, which is easier to calculate, was chosen for this study. The coefficients of the separation formula were set to match the two boundary lines shown in Figure 1 of Brinchmann et al. 2004. As a result, the boundary line (blue) between SFGs and CMP (Composite) was determined as follows:

$$\log\left(\frac{[OIII]5008}{H\beta}\right) = \frac{0.61}{\log\left(\frac{[NII]6585}{H\alpha}\right) - 0.07} + 1.30 \qquad (1)$$

CMP and AGN boundary line (red):

$$\log\left(\frac{[OIII]5008}{H\beta}\right) = \frac{0.61}{\log\left(\frac{[NII]6585}{H\alpha}\right) - 0.48} + 1.24 \qquad (2)$$

Using Table 2 from Kartaltepe et al. 2015, BPT diagrams were created for each redshift range, and the numbers of SFGs and AGNs were determined (Figure 2). From Figure 2, the number of SFGs and AGNs was converted into number density by calculating the volume of the spherical shell with a radius corresponding to the comoving radial distance (CRD) of the minimum and maximum redshift in each range. The survey volume was then calculated by multiplying the ratio of the survey area (COSMOS is 2 deg$^2$) to the global surface area (41,253 deg$^2$). The results are shown in Figure 3. SFGs tend to be distributed at low redshifts, while AGNs are distributed at high redshifts.

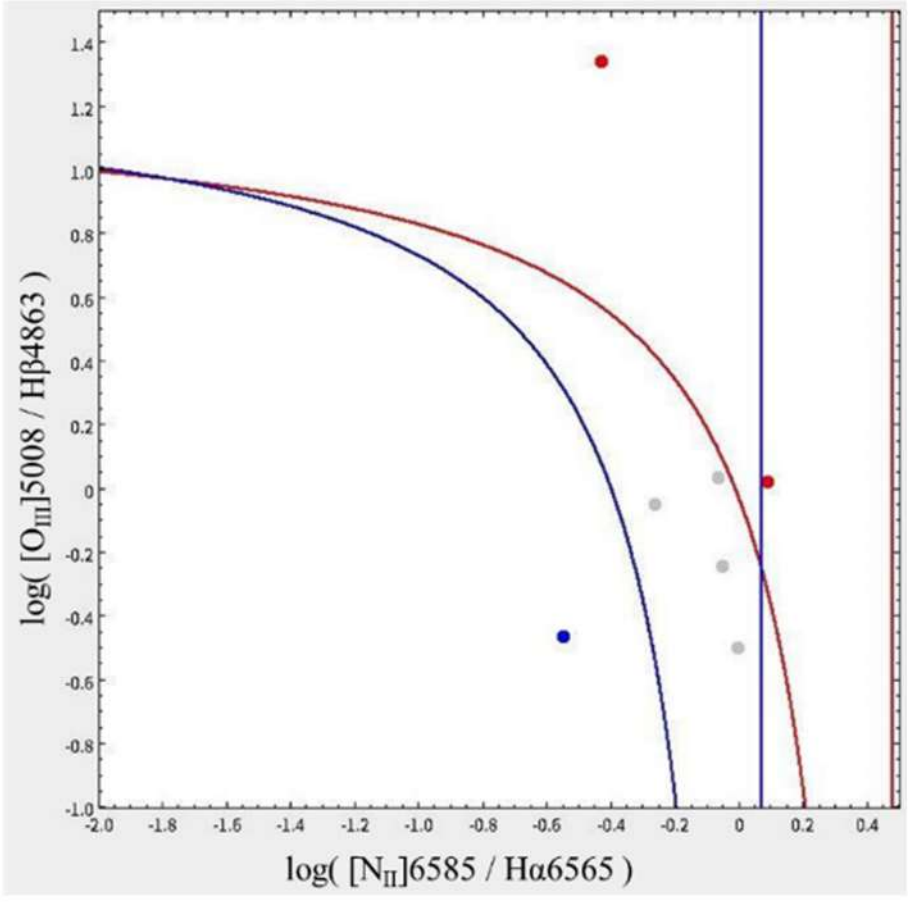

0.8 < z ≤ 1.0

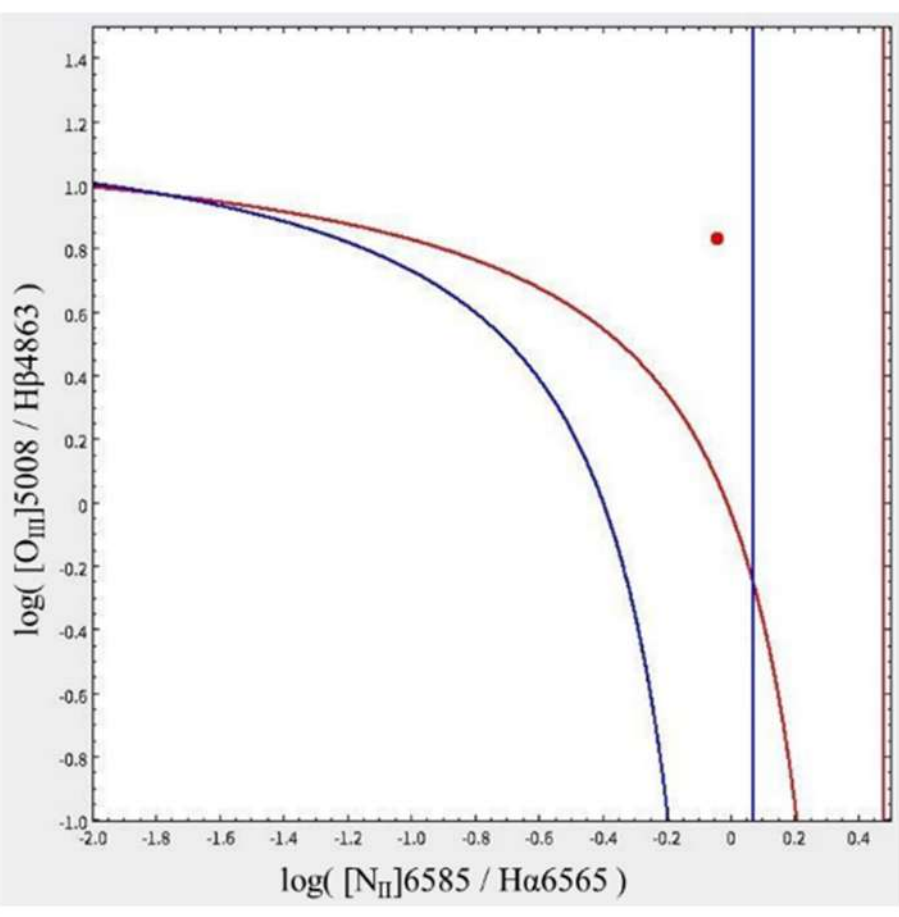

1.0 < z ≤ 1.2

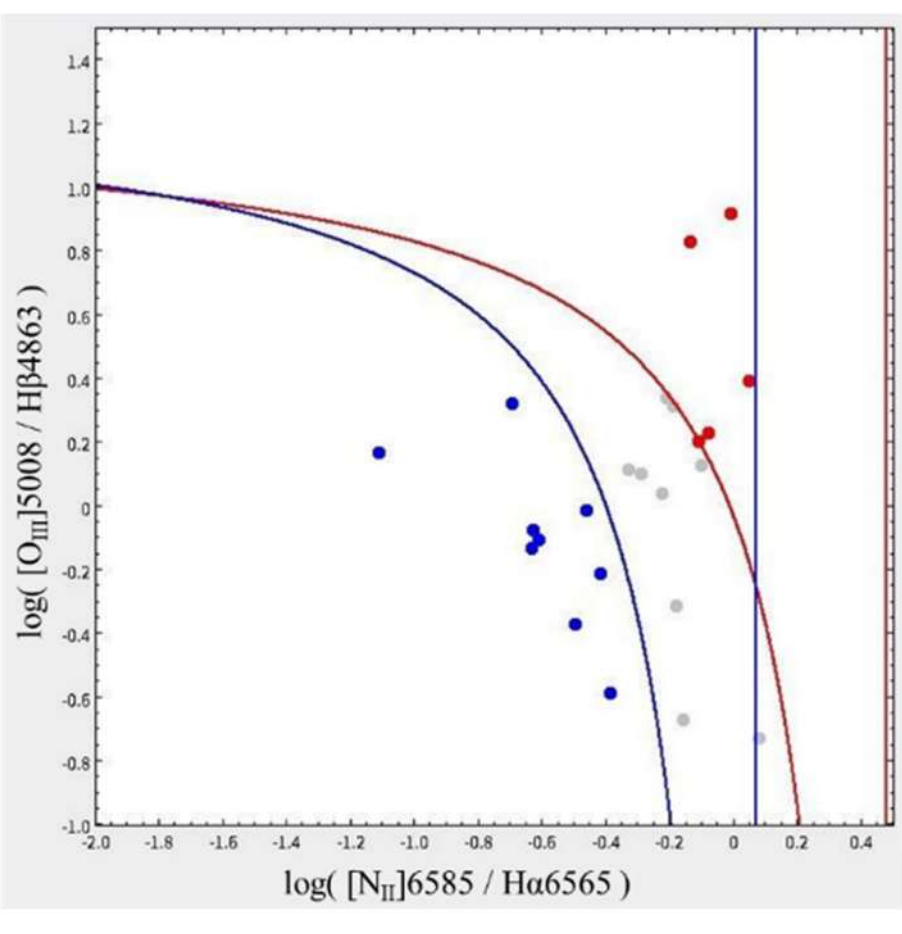

1.2 < z ≤ 1.4

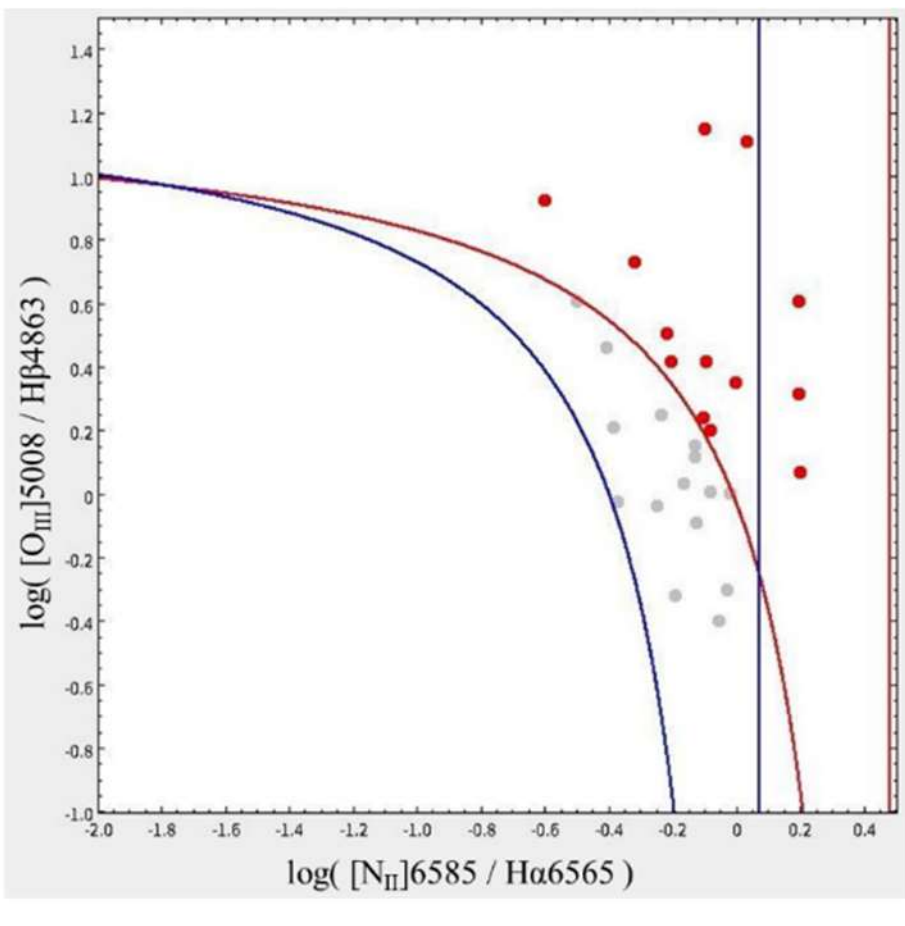

1.4 < z ≤ 1.6

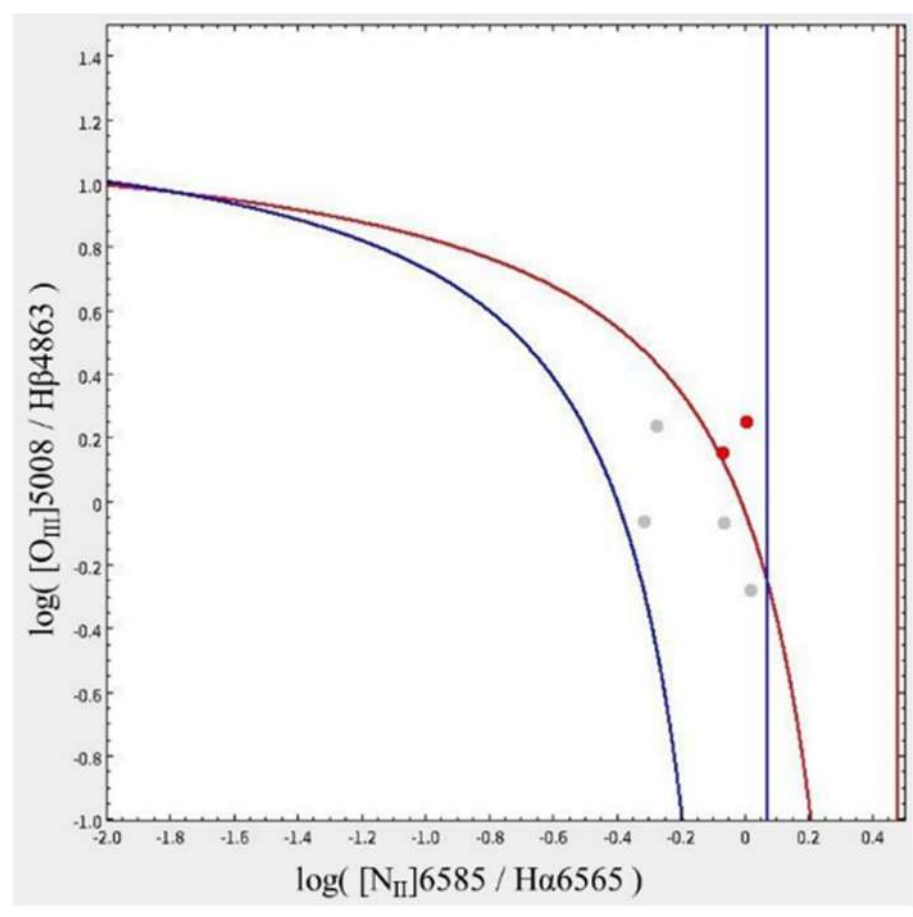

1.6 < z ≤ 1.8

**Figure 2.** BPT diagrams for five redshift ranges.
blue: SFG, gray: CMP, red: AGN. (Kartaltepe et al. 2015 Table 2)

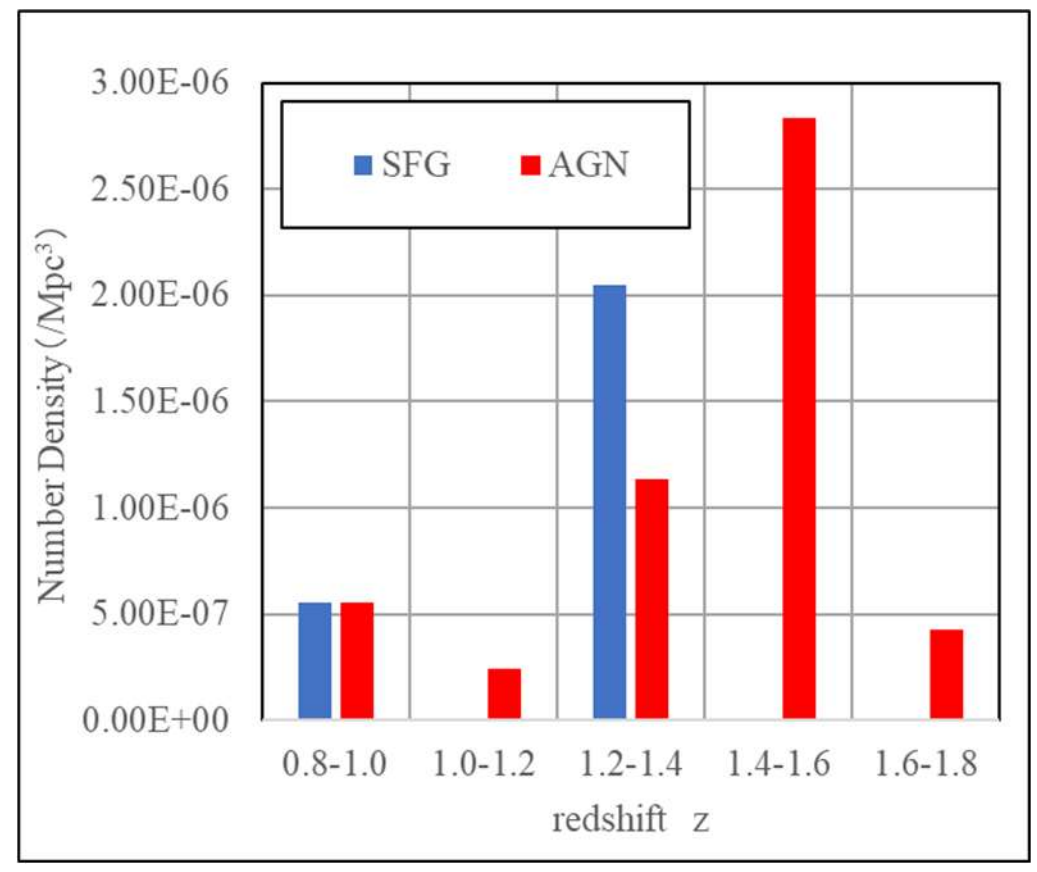

**Figure 3.** Dependence of SFG and AGN number density on redshift (Kartaltepe et al. 2015 Table 2)

*4.2 FMOS-COSMOS-CATALOG-2019 Analysis (Subaru-FMOS)*

In this catalog, the flux (erg $s^{-1}$ $cm^{-2}$) and flux error (erg $s^{-1}$ $cm^{-2}$) for four emission lines are recorded, so to understand the distribution trends of SFGs and AGNs depending on the S/N (flux/flux error), analyses were conducted for three cases: S/N ≥ 3, S/N ≥ 5, and S/N ≥ 10. Here, for S/N ≥ 5, the BPT diagrams for each redshift range (Figure 4) and the redshift dependence of number density (Figure 5) are shown (survey area is 1.7 deg²).

For S/N ≥ 5, SFGs are distributed in the three regions: $1.2 < z \leq 1.4$, $1.4 < z \leq 1.6$, and $1.6 < z \leq 1.8$, while AGNs are distributed only in the region $1.6 < z \leq 1.8$. The tendency for AGNs to exist in higher-z regions than SFGs was also observed in other S/N cases, but it became more remarkable as S/N improved and data reliability increased. From these results, it is considered that in the universe with $1.2 < z \leq 1.8$, AGNs are distributed in higher-z regions than SFGs.

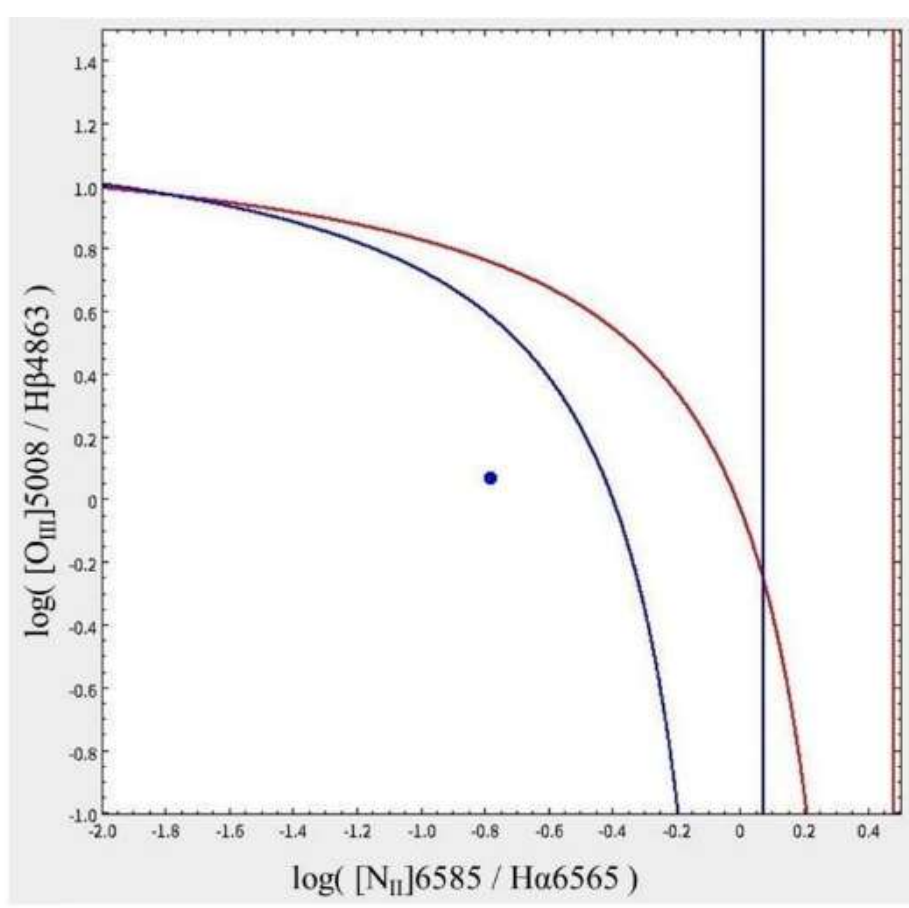

$1.2 < z \leq 1.4$

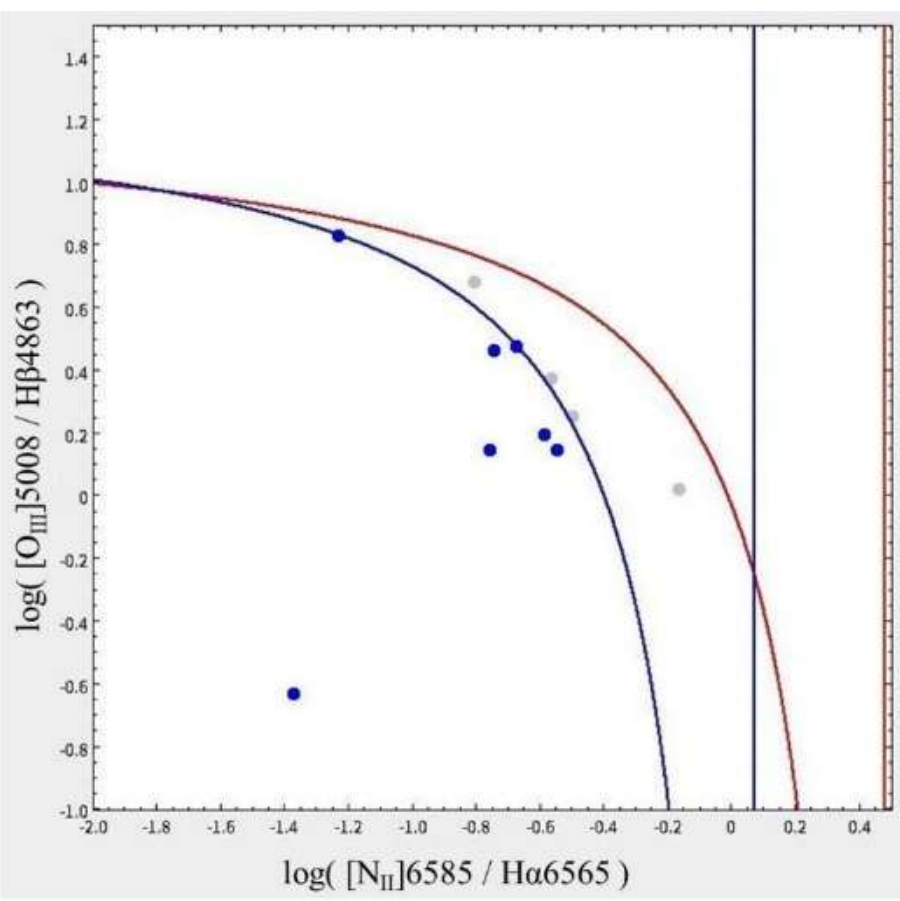

$1.4 < z \leq 1.6$

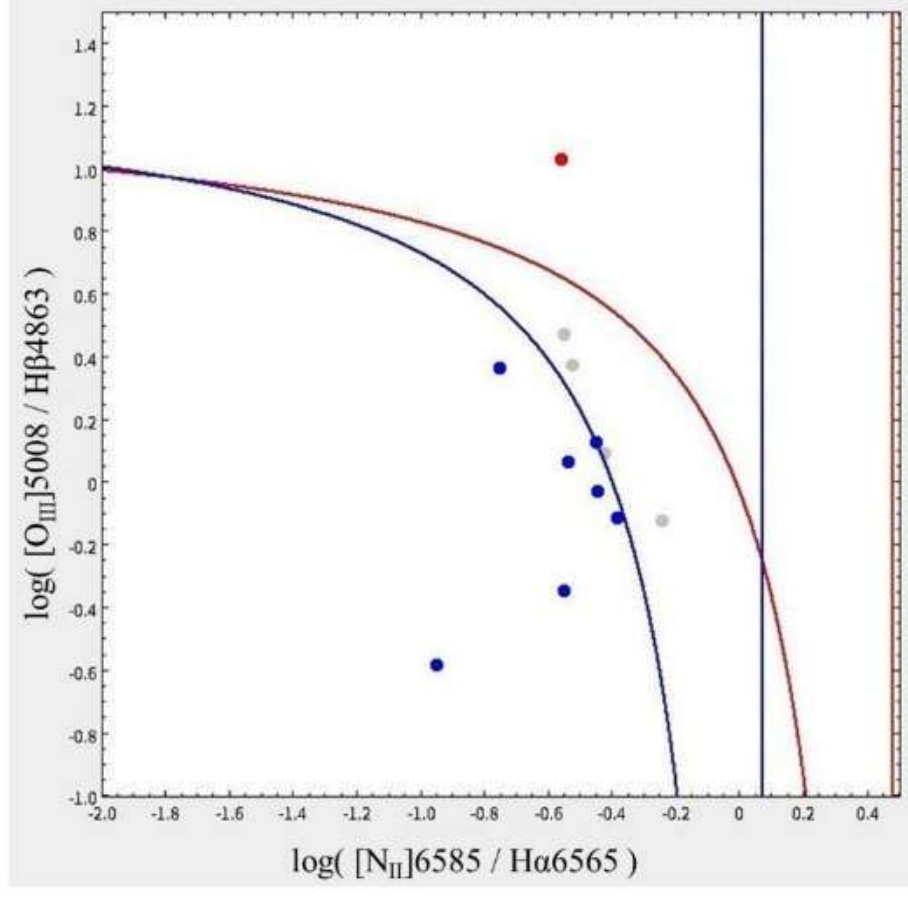

$1.6 < z \leq 1.8$

**Figure 4.** BPT diagrams for three redshift ranges with S/N ≥ 5 (FMOS-COSMOS-CATALOG-2019)

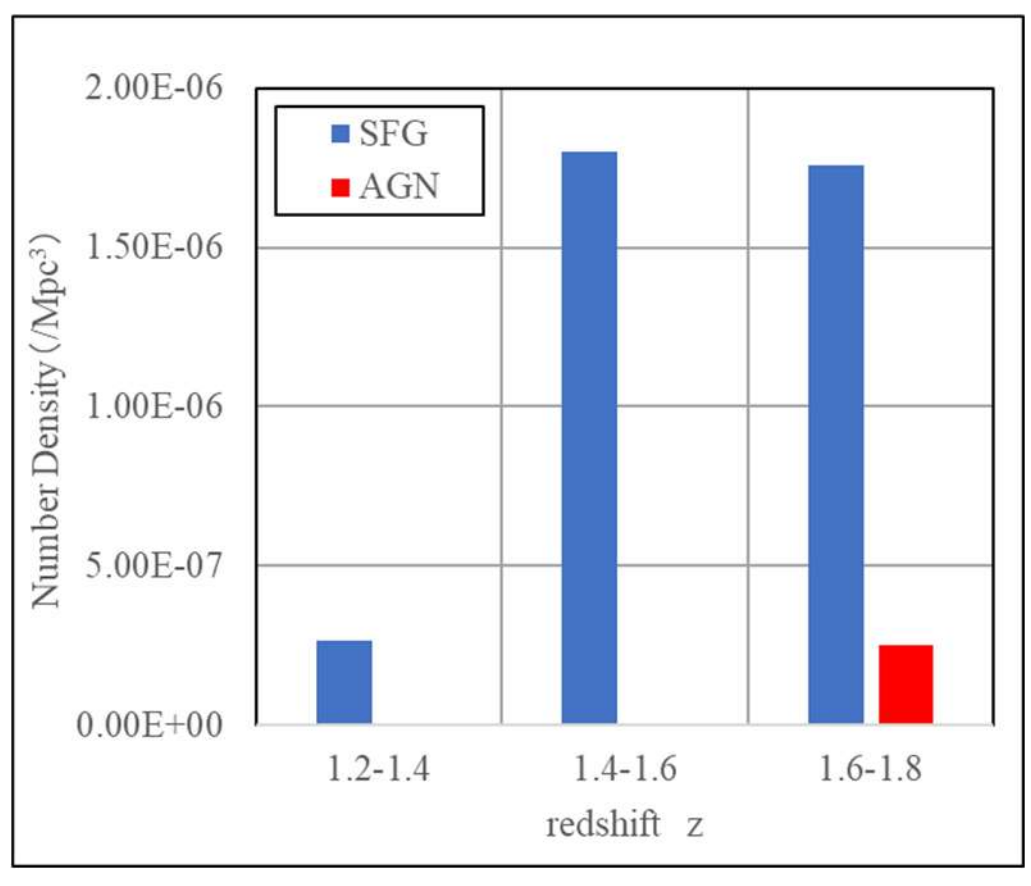

**Figure 5.** Redshift dependence of SFG and AGN number density with S/N ≥ 5 (FMOS-COSMOS-CATALOG-2019)

*4.3 Analysis of MOSFIRE-KBSS (KECK-MOSFIRE)*

From the 251 celestial objects listed in Table 1 of Steidel et al. 2014, which are in the redshift range of $2.0 < z \leq 2.6$, 168 objects with measurable fluxes for the four emission lines $[O_{III}]5008$, $[N_{II}]6585$, Hα, and Hβ were selected. These objects were divided into three redshift ranges ($2.0 < z \leq 2.2$, $2.2 < z \leq 2.4$, and $2.4 < z \leq 2.6$) in increments of 0.2, and BPT diagrams were created for each range to classify SFGs and AGNs (Figure 6). The redshift dependence of number density is shown in Figure 7 (survey area: 0.13 deg²).

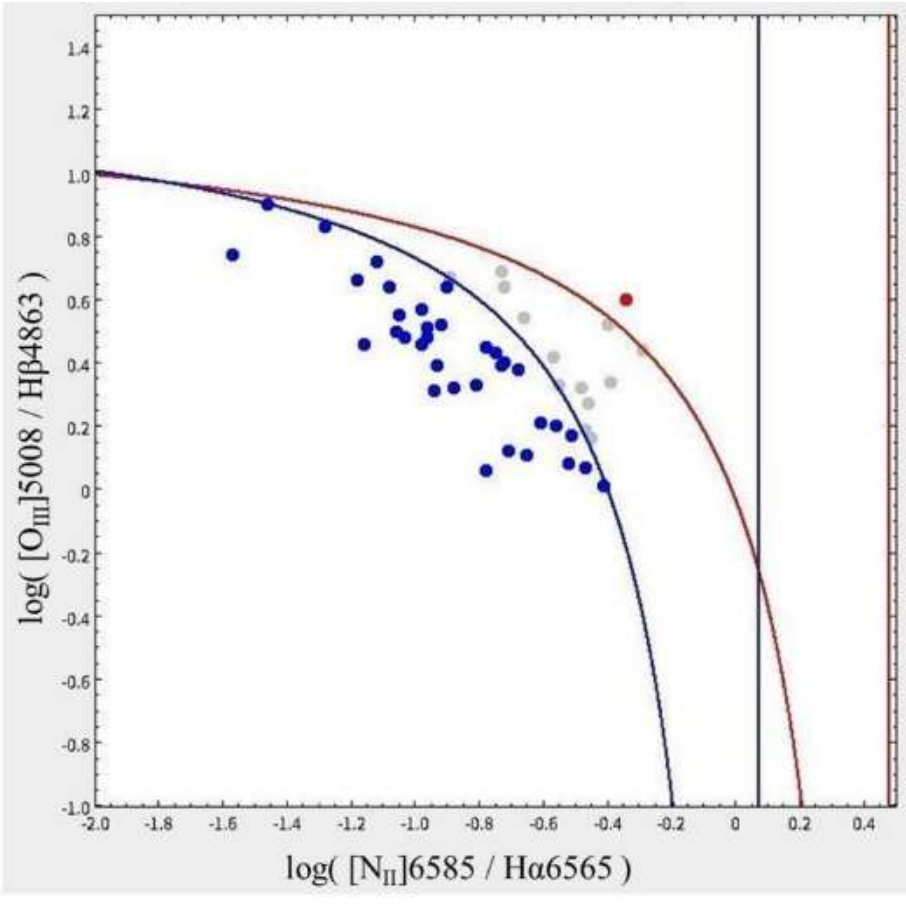

$2.0 < z \leq 2.2$

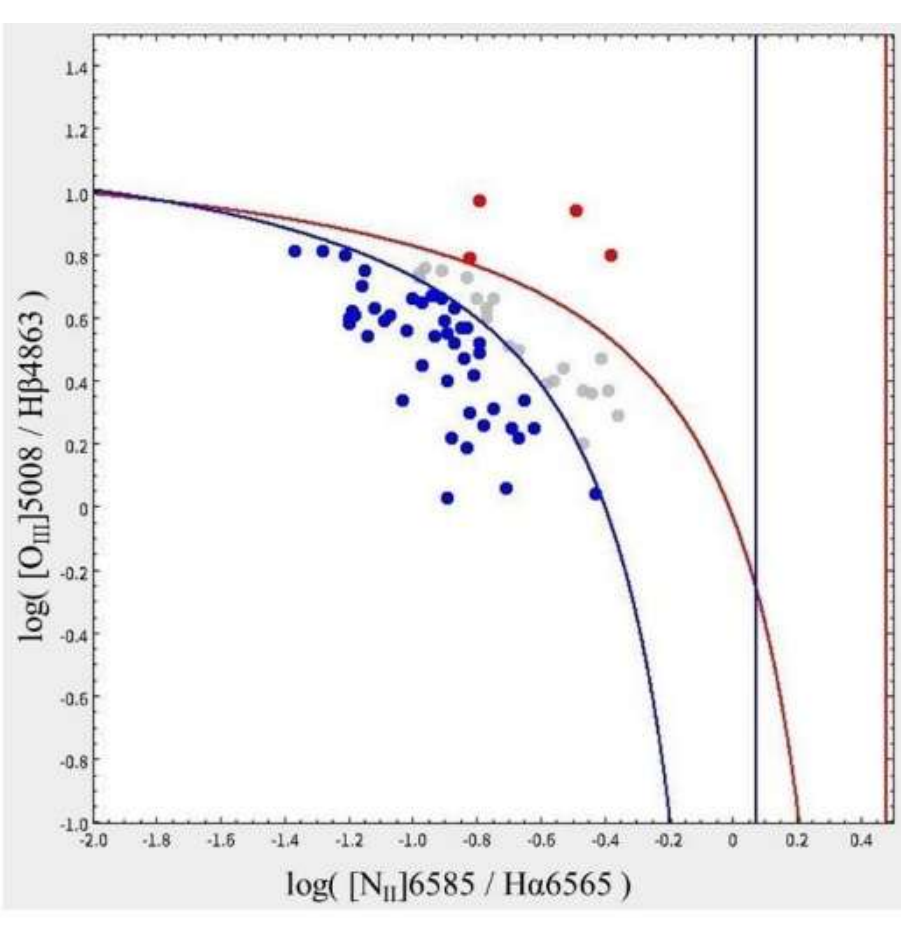

$2.2 < z \leq 2.4$

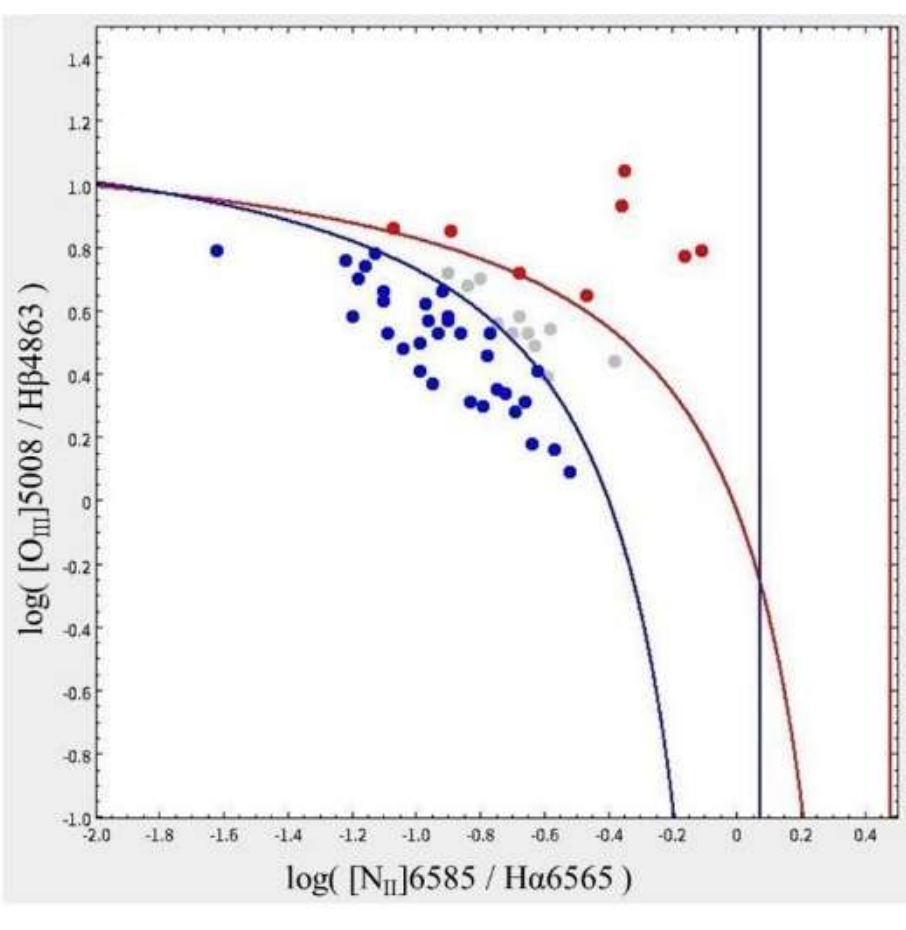

$2.4 < z \leq 2.6$

**Figure 6.** BPT diagrams for three redshift ranges (MOSFIRE-KBSS)

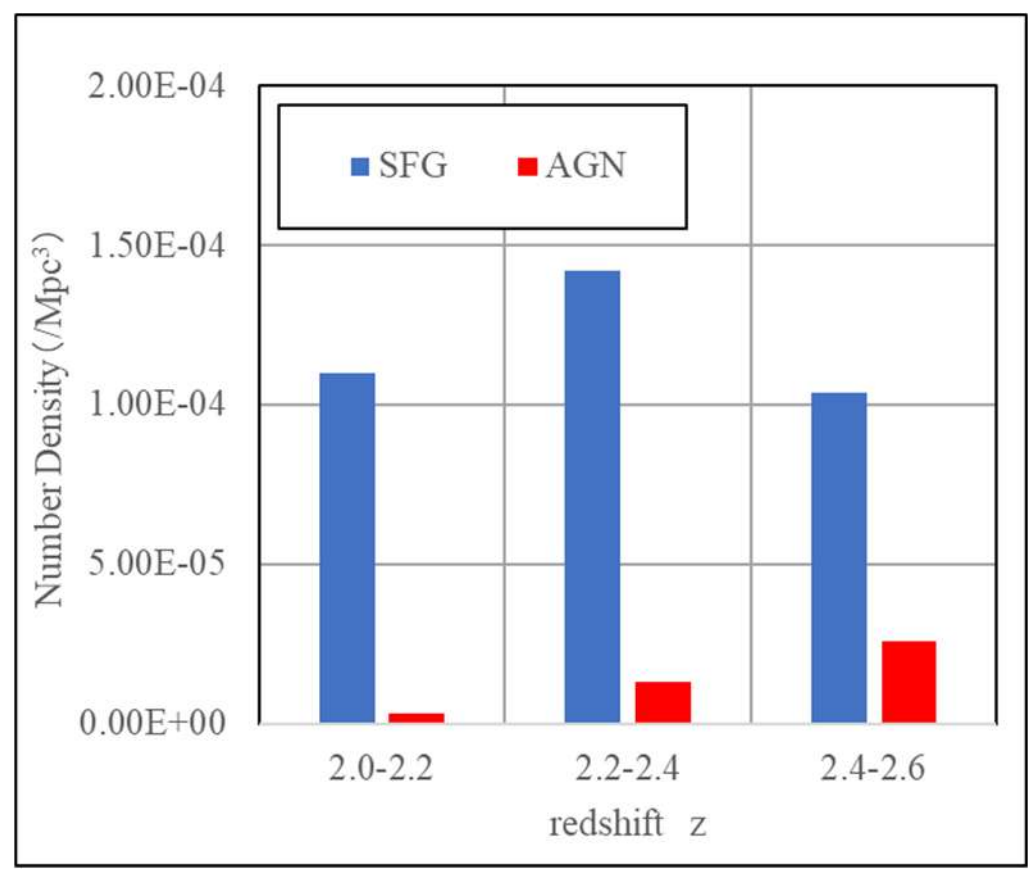

**Figure 7.** Redshift Dependence of Number Density for SFGs and AGNs (MOSFIRE-KBSS)

From Figure 7, it was observed that SFGs were distributed relatively evenly across the three redshift ranges, while AGNs showed an increase in number density as the redshift increased. The distribution peak for SFGs was in the range $2.2 < z \leq 2.4$, while for AGNs, the peak was in the range $2.4 < z \leq 2.6$ and beyond. This indicates that in the universe within the range of $2.0 < z \leq 2.6$, SFGs are distributed in lower redshift regions, while AGNs are distributed in higher redshift regions.

*4.4 Analysis of MOSFIRE-MOSDEF (KECK-MOSFIRE)*

This catalog contains 22 types of information, including the flux values (FLUX, in units of erg s$^{-1}$ cm$^{-2}$) and flux errors (FLUX_ERR, in units of erg s$^{-1}$ cm$^{-2}$) for 16 emission lines ranging from [O$_{II}$]3727 to [S$_{II}$]6733. To understand the distribution trends of SFGs and AGNs based on the signal-to-noise ratio (S/N = FLUX/FLUX_ERR), the flux and flux error ratios for the four emission lines required for BPT diagram analysis ([O$_{III}$]5008, [N$_{II}$]6585, Hα, and Hβ) were calculated. Analyses were conducted for four different S/N thresholds: S/N ≥ 3, S/N ≥ 5, S/N ≥ 10, and S/N ≥ 15. Here, the analysis for S/N ≥ 10 is presented, including BPT diagrams for each redshift range (Figure 8) and the redshift dependence of number density (Figure 9). The survey area was 0.13 deg².

For S/N ≥ 10, SFGs were distributed across five redshift ranges: $1.2 < z \leq 1.4$, $1.4 < z \leq 1.6$, $1.6 < z \leq 1.8$, $2.0 < z \leq 2.2$, and $2.2 < z \leq 2.4$. AGNs, on the other hand, were only distributed in the ranges $2.2 < z \leq 2.4$ and $2.4 < z \leq 2.6$ (Figure 8).

For S/N ≥ 15, SFGs were distributed across four redshift ranges: $1.2 < z \leq 1.4$, $1.4 < z \leq 1.6$, $1.6 < z \leq 1.8$, and $2.2 < z \leq 2.4$, while AGNs were only present in the range $2.4 < z \leq 2.6$. This suggests that AGNs tend to exist in higher redshift regions compared to SFGs. On the other hand, at lower S/N thresholds (S/N ≥ 3 and S/N ≥ 5), no clear distribution trends were observed, likely due to the influence of noise. From the above, it can be concluded that in the universe within the range of $1.2 < z \leq 2.6$, AGNs are distributed in higher redshift regions compared to SFGs.

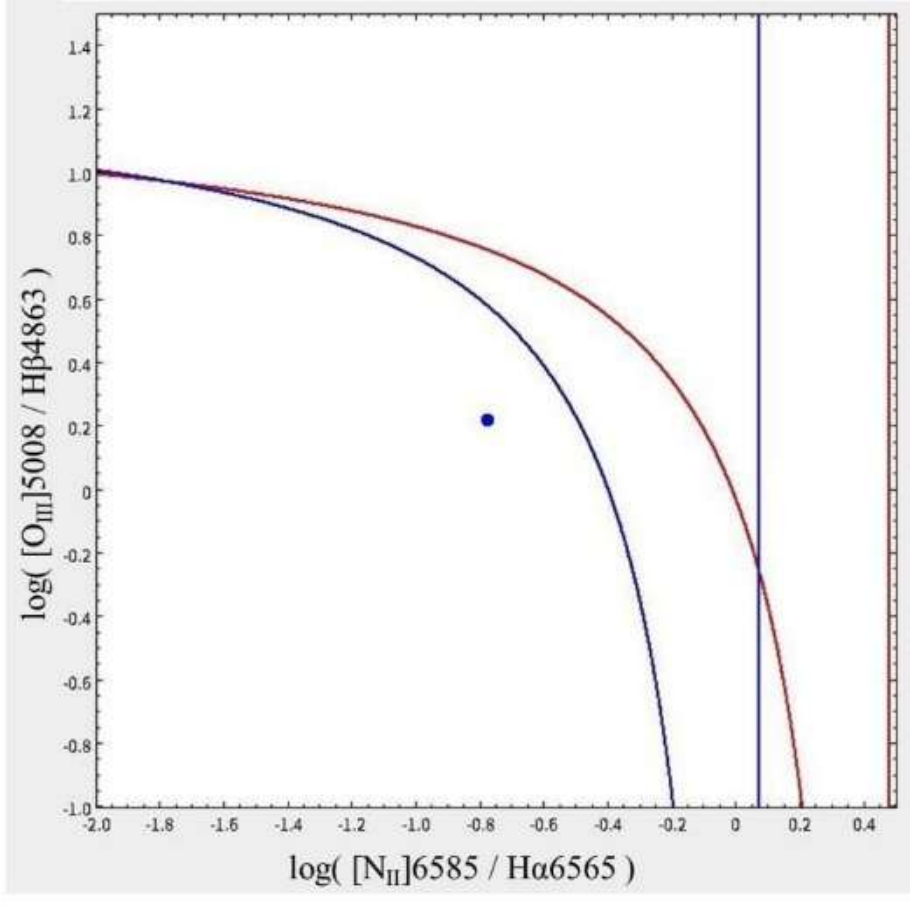

$1.2 < z \leq 1.4$

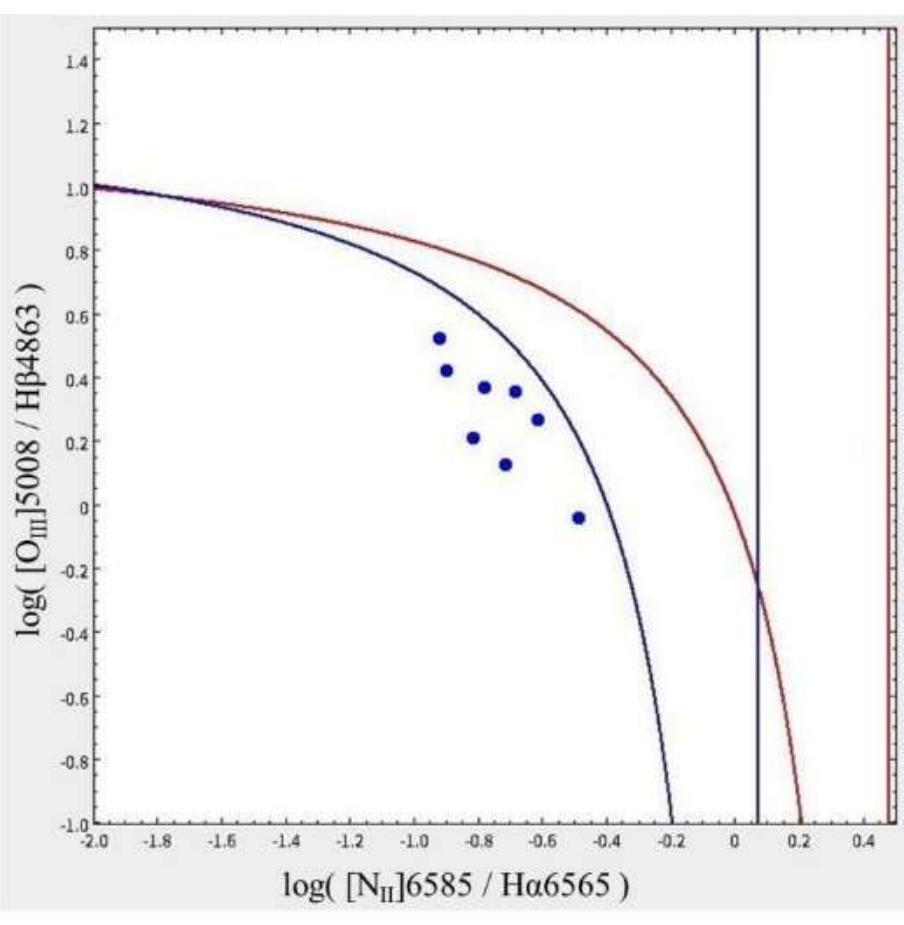

$1.4 < z \leq 1.6$

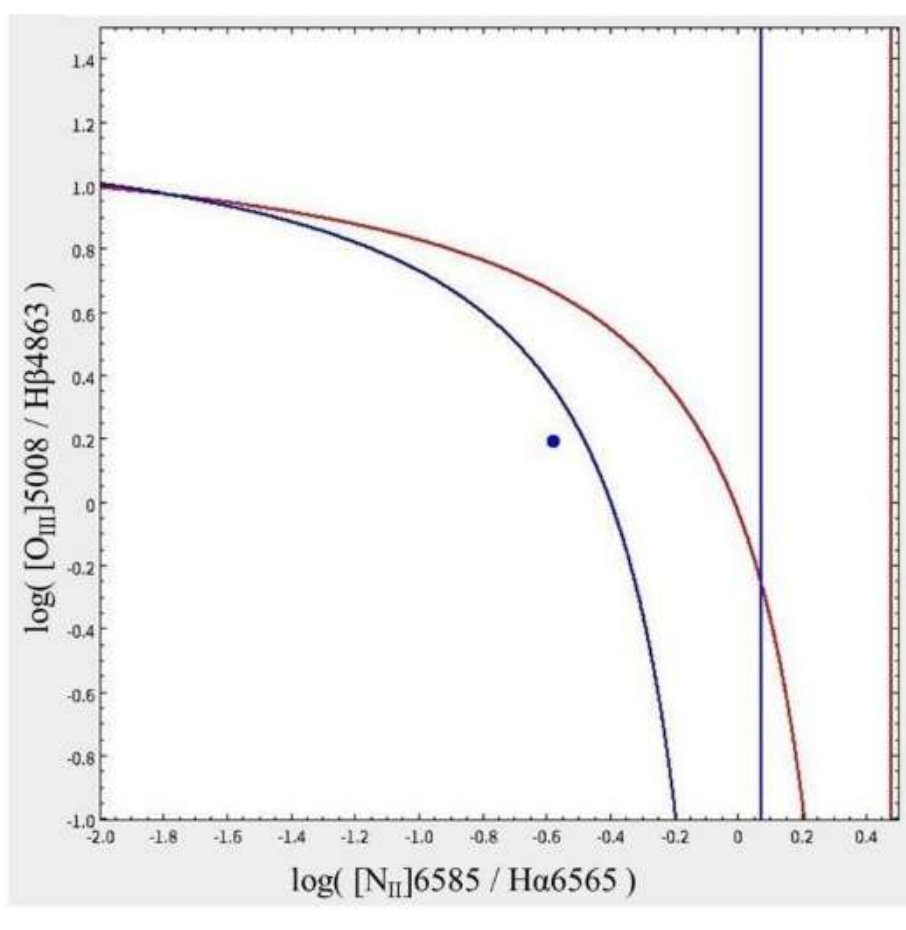

$1.6 < z \leq 1.8$

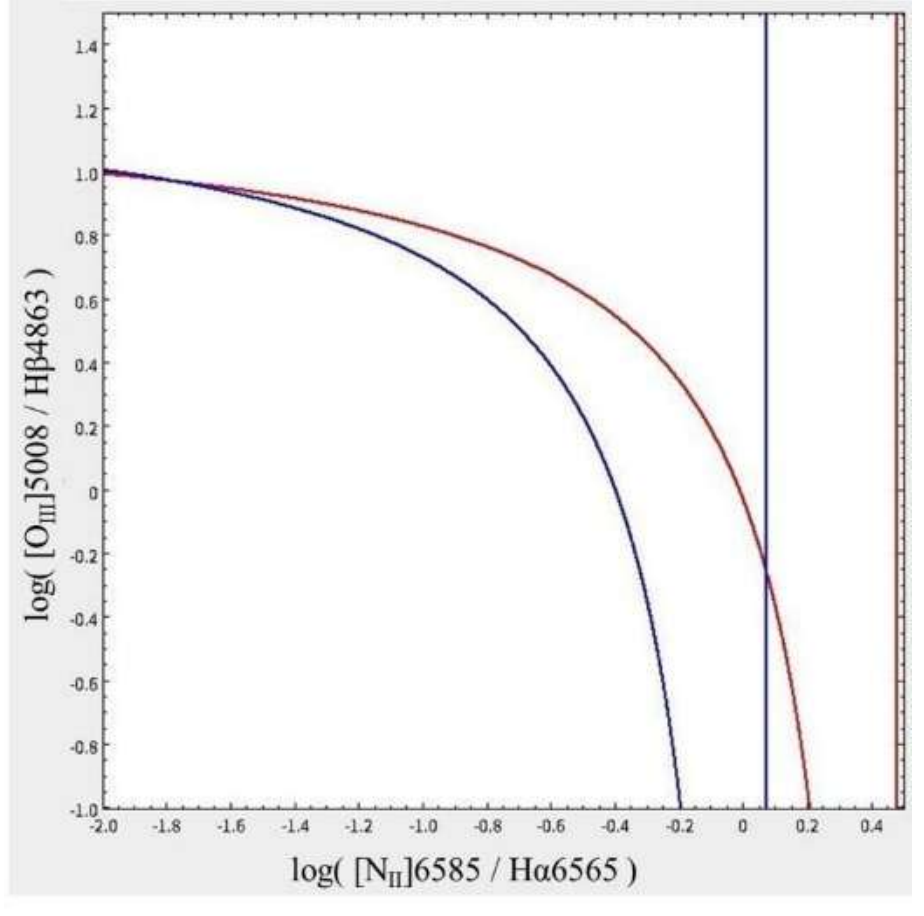

$1.8 < z \leq 2.0$

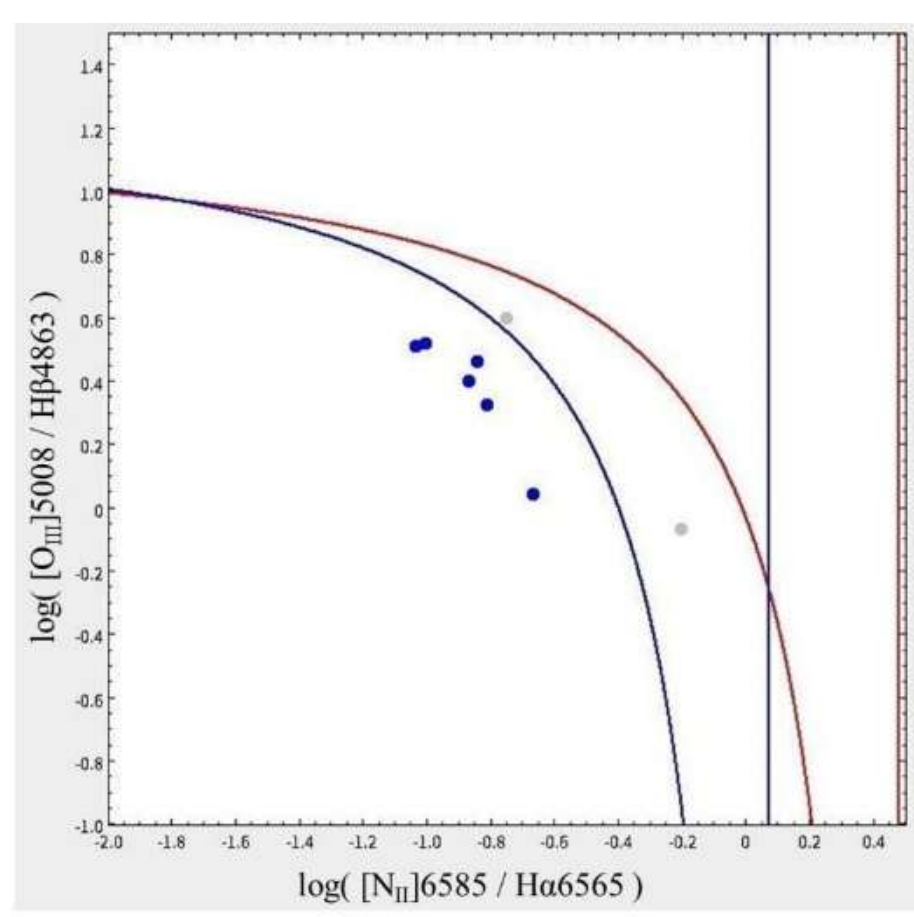

$2.0 < z \leq 2.2$

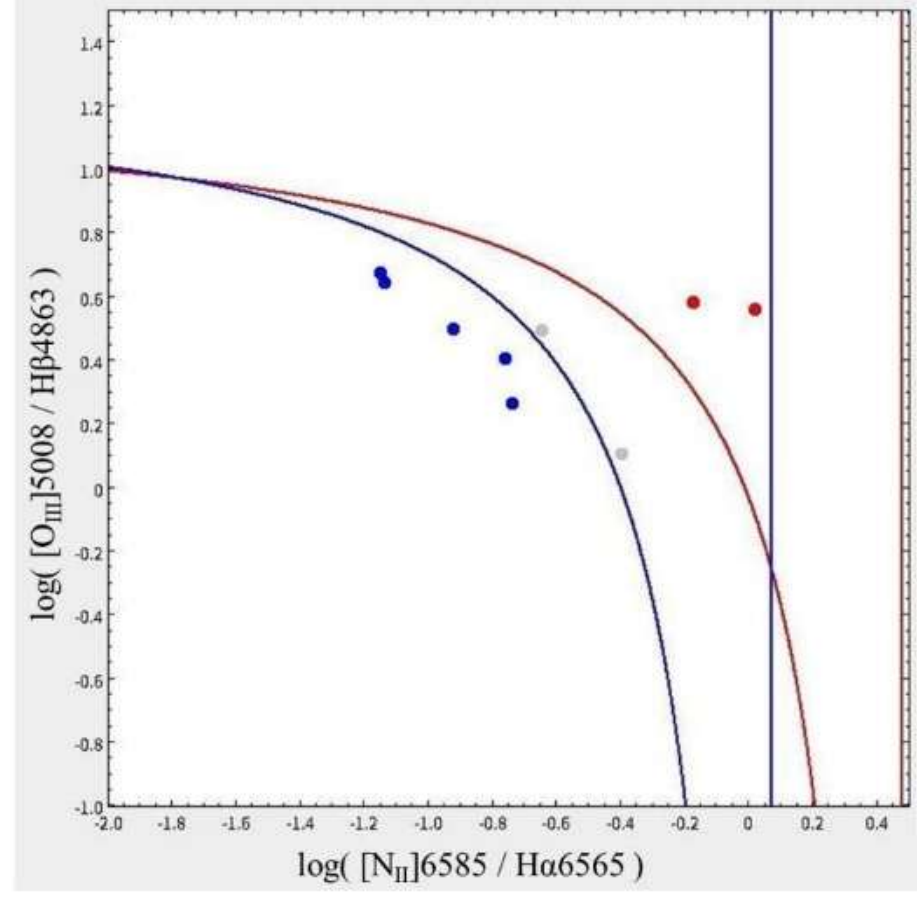

$2.2 < z \leq 2.4$

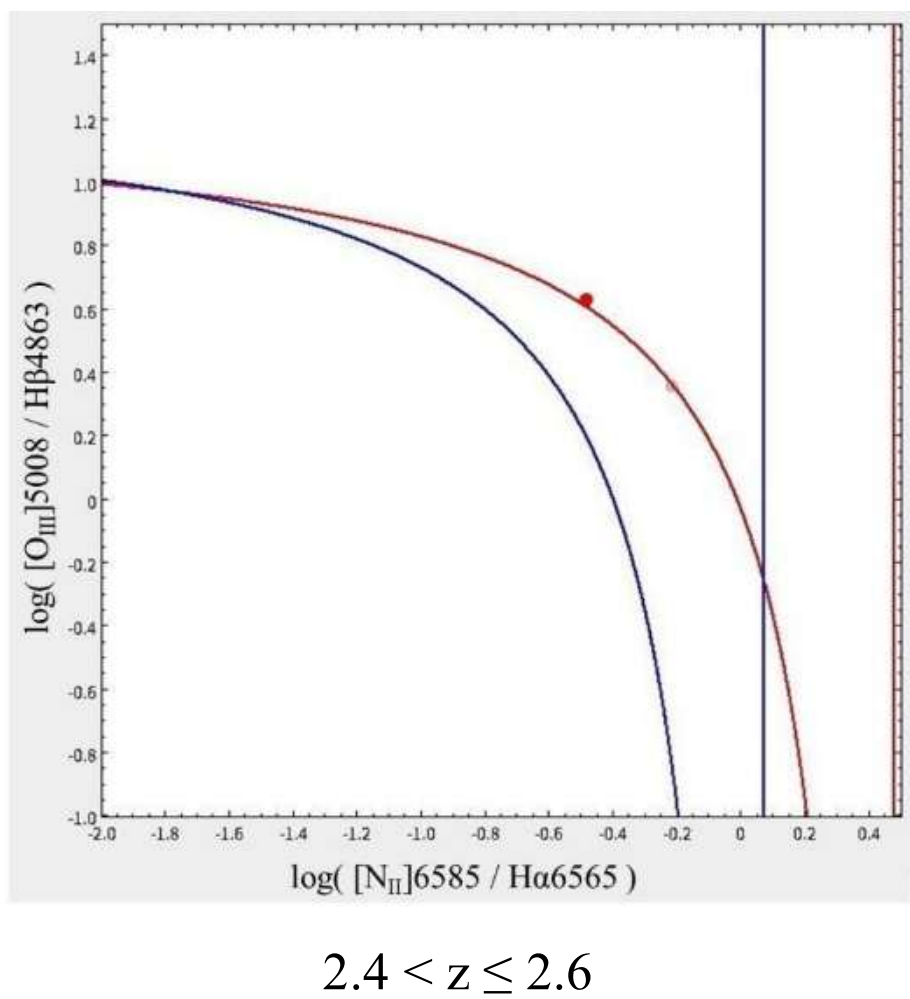

$2.4 < z \leq 2.6$

**Figure 8.** BPT Diagrams of seven redshift ranges for S/N ≥ 10 (MOSFIRE-MOSDEF)

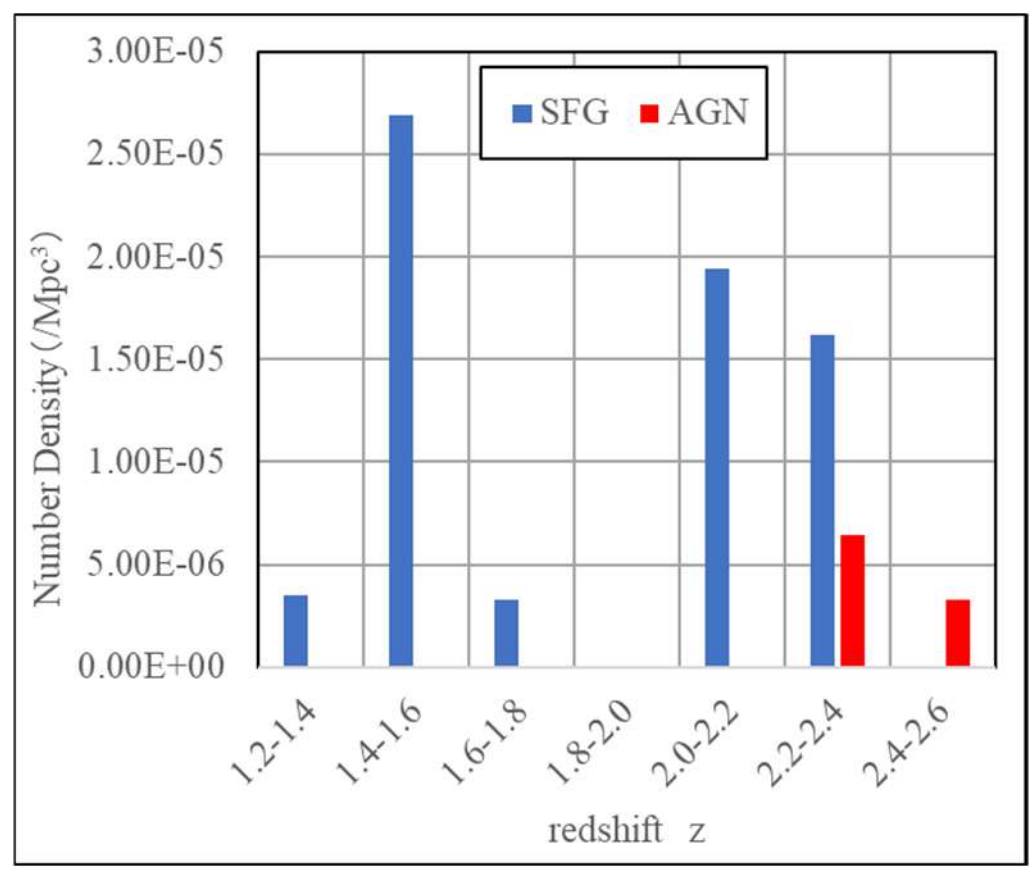

**Figure 9.** Redshift Dependence of Number Density of SFGs and AGNs for S/N ≥ 10 (MOSFIRE-MOSDEF)

*4.5 Analysis of High-Redshift Region Measurement Papers Using JWST/NIRSpec*

Figure 3 of Sanders et al. 2023 includes BPT diagrams based on various emission line luminosity ratios for 164 celestial objects in the redshift range of z=2.0–9.3. This study focuses on the [$N_{II}$]6585 BPT diagram (vertical axis: [$O_{III}$]5008/Hβ, horizontal axis: [$N_{II}$]6585/Hα) and the [$S_{II}$]6718 BPT diagram (vertical axis: [$O_{III}$]5008/Hβ, horizontal axis: [$S_{II}$]6718/Hα) to analyze the redshift dependence of SFGs and AGNs across various redshift ranges.

In both the [$N_{II}$]6585 BPT diagram and the [$S_{II}$]6718 BPT diagram, it was observed that in the redshift range $2.7 \leq z < 4.0$, SFGs shifted to the upper right compared to the range $2.0 \leq z < 2.7$, moving closer to the SFG/AGN boundary. As the redshift increased further to ranges such as $4.0 \leq z < 5.0$ and $5.0 \leq z < 6.5$, the trend of shifting to the upper right became ambiguous, and instead, the galaxies appeared to move upward. This is likely because, as galaxies exist farther away, the [$O_{III}$]5008 emission line with good S/N can still be measured, but the luminosity of [$N_{II}$]6585 and [$S_{II}$]6718 approaches the detection limit. In any case, as the redshift increases, the proportion of galaxies distributed in the AGN region increases, and they tend to be located farther from the SFG/AGN boundary line. Although the redshift dependence of SFGs and AGNs is qualitative, it is summarized in Figure 10. Specifically, at low redshifts, SFG > AGN, while at high redshifts, SFG < AGN.

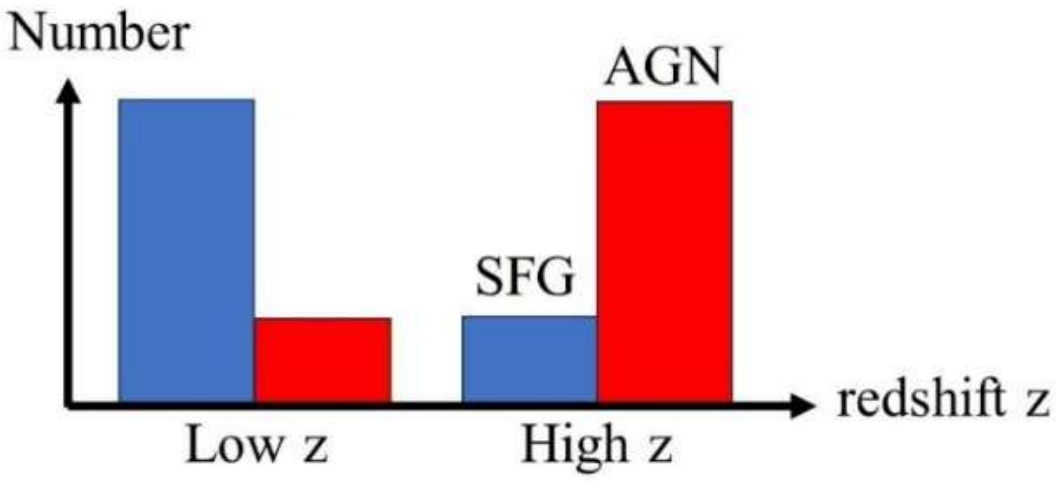

**Figure 10.** Qualitative dependence of SFG and AGN on redshift z, obtained from the BPT diagram for [$N_{II}$]6585 and [$S_{II}$]6718 from Figure 3 of Sandars et al. 2023.

*4.6 Analysis of High-Redshift Region Measurement Papers Using JWST/MIRI*

Yang et al. 2023 observed the Extended Groth Strip (EGS) region located between the constellations Ursa Major and Boötes using JWST's MIRI instrument. The results were analyzed using SED fitting/modeling with CIGALE, classifying galaxies into SFGs (referred to as SF in the paper), mixed, and AGNs.

From Figure 3 of Yang et al. 2023, the number of SFGs and AGNs in each redshift range was extracted. Using the survey area (9 arcmin²), the survey volume was calculated, and the number density of SFGs and AGNs was determined. The redshift dependence of the number density was summarized in Figure 11.

From Figure 11, the trends in the number density of SFGs and AGNs with respect to redshift can be observed. SFGs are more numerous than AGNs at low redshifts but decrease as redshift increases. In contrast, AGNs remain relatively constant regardless of redshift, and beyond z~3, AGNs become more numerous than SFGs. This indicates that SFGs are more prevalent at low redshifts, while AGNs dominate at higher redshifts.

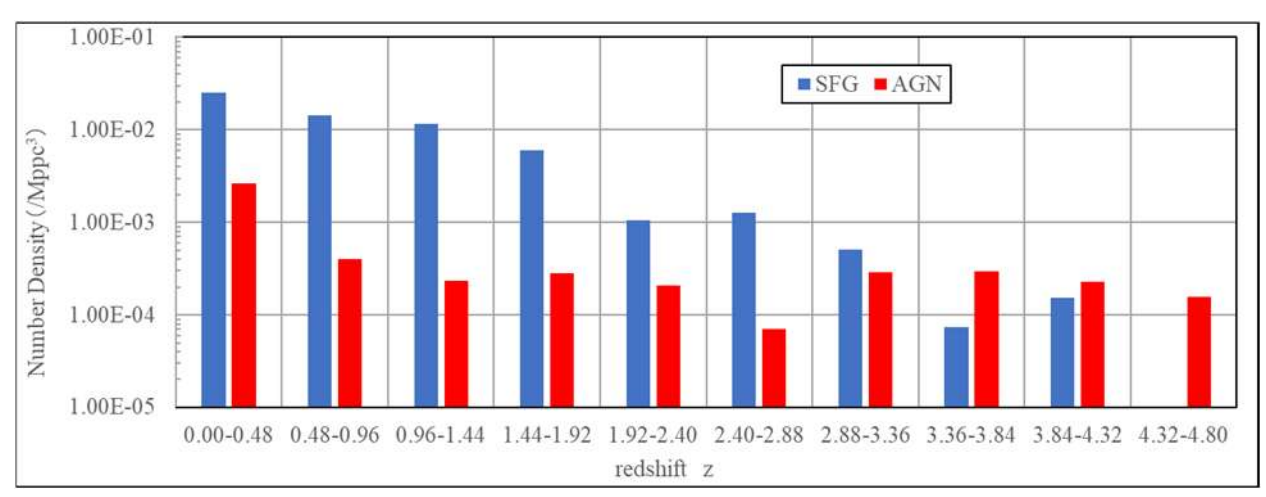

**Figure 11.** Dependence of SFG and AGN number density on redshift z, obtained from the Figure 3 of Yang et al. 2023.

## 5. DISCUSSION

From the redshift dependence of the number density of star-forming galaxies (SFGs) and active galactic nuclei (AGNs) observed using Subaru-FMOS, KECK-MOSFIRE, and JWST, the overall trend of the survey is summarized in Figure 12.

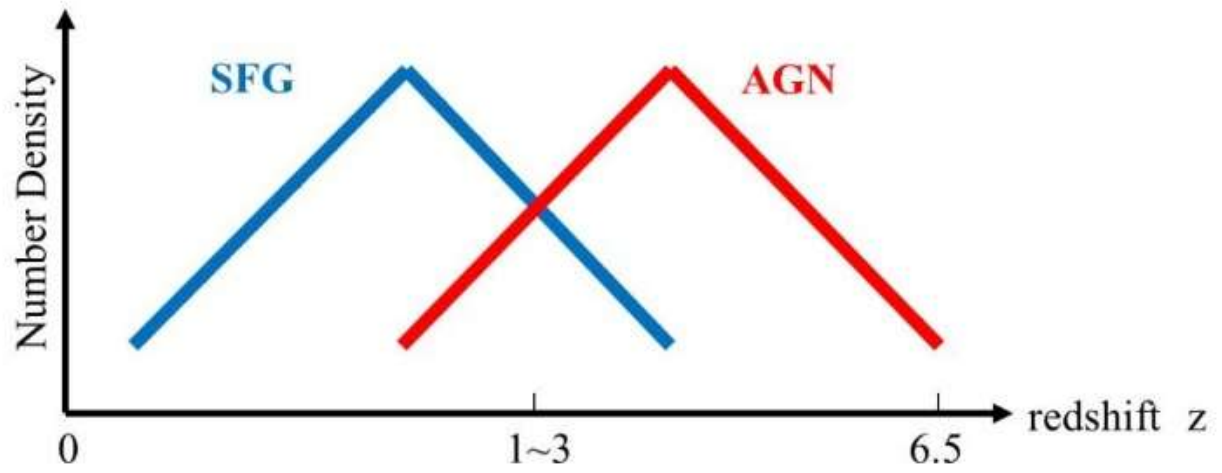

**Figure 12.** Redshift dependence of the number density of SFGs and AGNs observed using Subaru-FMOS, KECK-MOSFIRE, and JWST (Overall survey trend).

From Figure 12, the following trends in density evolution can be observed:

**Trend (1):** "SFGs are distributed at low redshifts, while AGNs are distributed at high redshifts ($z < 6.5$)."

**Trend (2):** "The number density of SFGs and AGNs reverses between z=1 and z=3."

As mentioned earlier, the mechanisms driving the luminosity and density evolution of galaxies are thought to involve:

**Factor (A):** "SFGs merge and transform into AGNs."

**Factor (B):** "AGN activity decreases."

For the mechanism of Factor (B), gas depletion and ADAF are considered. In this study, the focus was on gas depletion, and the relationship between the mass accretion rate of gas and AGN luminosity was confirmed through simple calculations.

If the gas with mass M near the center of a galaxy undergoes mass accretion onto the accretion disk due to the gravity of the supermassive black hole (SMBH), and its energy is emitted as electromagnetic radiation (luminosity $L_Q$), then the mass accretion time constant is τ, the efficiency of converting mass into electromagnetic energy is η (approximately 0.1), the light velocity is c, and time is t. Assuming A as a constant, the relationship can be expressed as:

$$\frac{dM}{dt} = -\frac{M}{\tau},\ M = A\exp\left(-\frac{t}{\tau}\right).$$

$$L_Q = \eta\left(-\frac{dM}{dt}\right)c^2 = \frac{\eta c^2 A}{\tau}\exp\left(-\frac{t}{\tau}\right)$$

$$\log(L_Q) = \log(e)\ln\left(\frac{\eta c^2 A}{\tau}\right) - \frac{t}{\tau/\log(e)} \qquad (3)$$

By analyzing the median luminosity values against redshift (z) from Kartaltepe et al. 2015 Figure 1 and converting the horizontal axis from redshift to cosmic age, the graph shown in Figure 13 was obtained. In Figure 13, the luminosity of galaxies monotonically decreases with increasing cosmic age. The average time constant for luminosity decline, $\tau_m$, between 1.6 billion years and 13.8 billion years, was calculated using Equation 3, resulting in $\tau_m \approx 1.5$ billion years.

At a certain point after the Big Bang, if all galaxies were born simultaneously, then 13.8 billion years later after Big Bang, at the present time, all galaxies observed at various redshifts (z) would have the same elapsed time since their birth. However, due to the finite velocity of electromagnetic wave propagation, galaxies with higher redshifts (z) appear as they were in the early universe.

In other words,

**Trend (3):** "Galaxies that were just born are bright, and they become dimmer over time."

This indicates a trend of luminosity evolution.

Additionally, Figure 13 consists of two decay lines with different slopes, which can be divided into the "period of rapid decay ($\tau_1 = 0.9$ billion years)" during the cosmic age of 1.6 billion to 2.8 billion years, and the "period of slow decay ($\tau_2 = 1.7$ billion years)" from 4.4 billion years to the present. As AGNs transform into SFGs and the number of SFGs increases, major mergers and minor mergers lead to the merging of SFGs, and Factor (A) begins to account for a certain proportion of the process. Both the "period of rapid decay" and the "period of slow decay" show a monotonic decrease in galaxy luminosity, indicating that Factor (B) is more dominant than Factor (A). However, during the "period of rapid decay," Factor (A) is less significant, resulting in a smaller τ. As the "period of slow decay" begins, the increase in SFGs leads to a rise in Factor (A), which in turn increases τ.

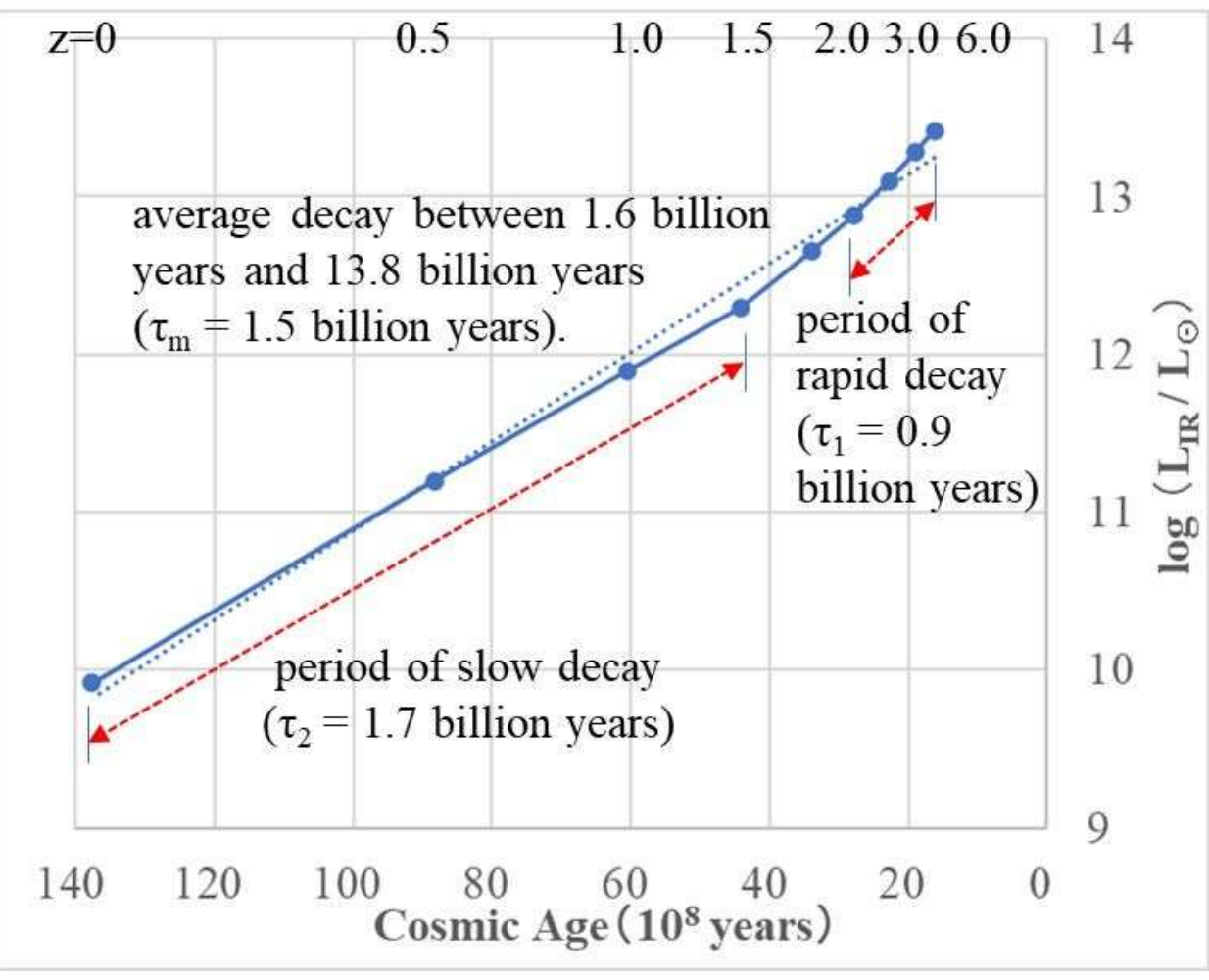

**Figure 13.** Dependence of galaxy IR luminosity on cosmic age. Time constants are calculated using equation (3).

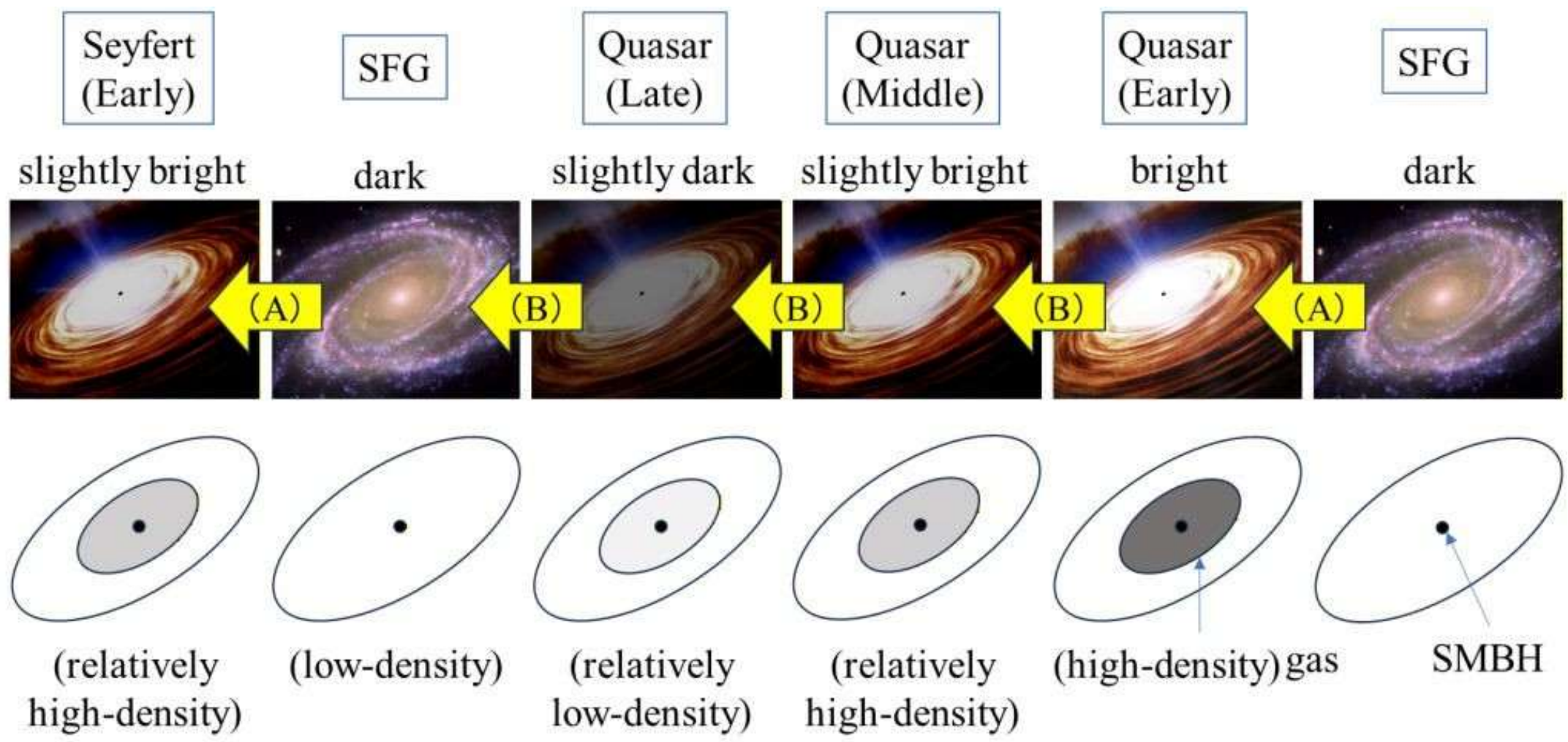

**Figure 14.** Model illustrating the transformation of SFGs into quasars, Seyfert galaxies, and LINERs.

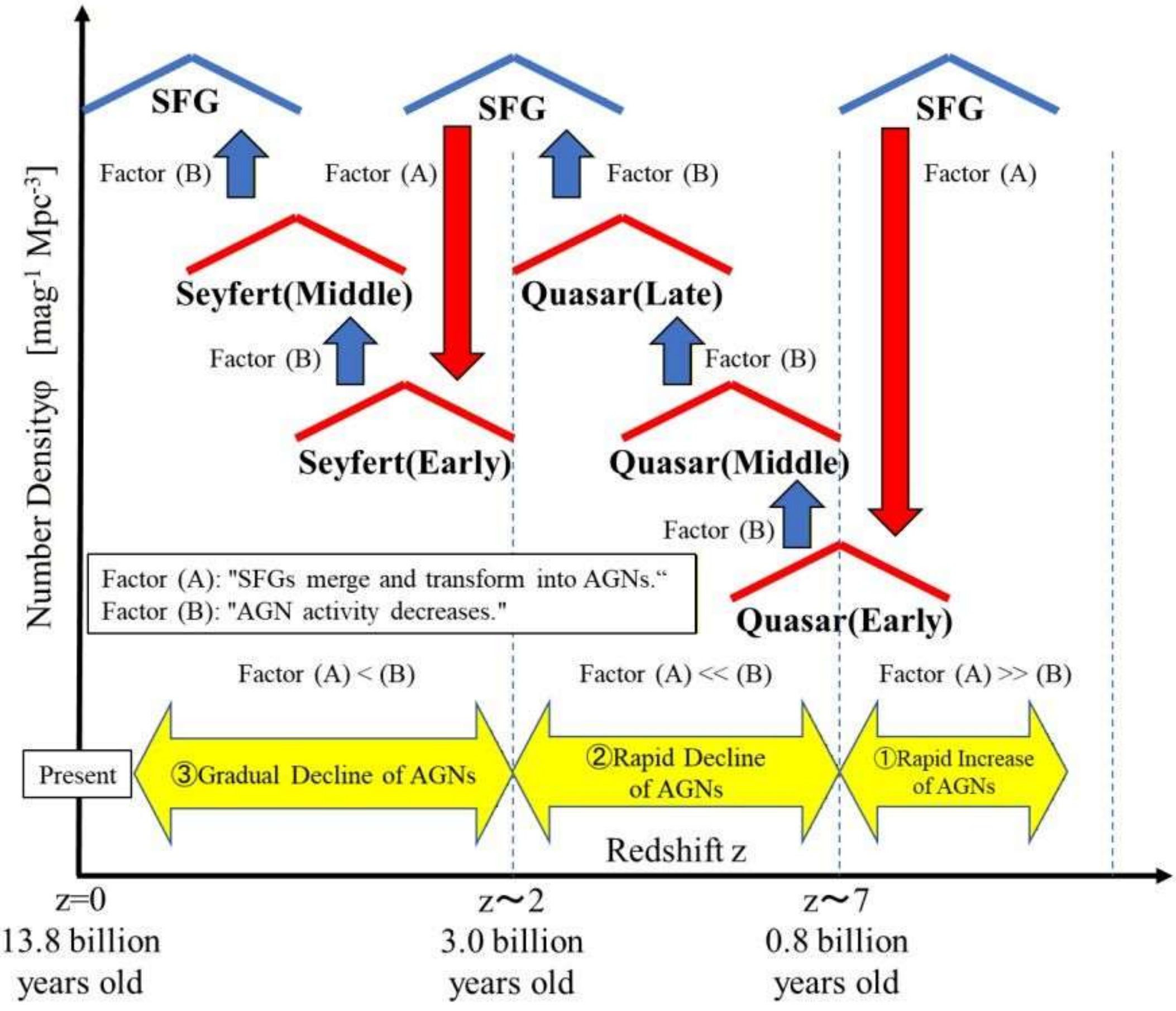

**Figure 15.** Schematic graph of the model showing the number density of AGNs and SFGs as a function of redshift (z).

By simultaneously satisfying Trend (1) and Trend (3), and incorporating the processes in the early universe where quasars are formed from SFGs due to Factor (A), the process of AGN activity decreasing, and the process in the nearby universe where AGNs such as Seyfert galaxies are reformed from SFGs, the model shown in Figure 14 is obtained. When this model is schematically graphed with the vertical axis representing number density and the horizontal axis representing redshift (z), it results in Figure 15.

### *5.1 Phase ①: Rapid Increase of AGNs (Factor (A) >> Factor (B))*

Phase ①, the rapid increase of AGNs, corresponds to the early universe, from the birth of the universe ($z=\infty$, age 0) to the formation of the first SMBHs and massive quasars ($z=7$, 800 million years after the Big Bang). During this period, massive stars with masses around 1000 $M_\odot$ were likely formed, leading to the creation of star clusters in regions of high number density. These clusters underwent "runaway mergers," resulting in the formation of supermassive stars. These stars quickly underwent supernova explosions, leaving behind intermediate-mass black holes (IMBHs) with masses around 100 $M_\odot$. Simulations investigating how IMBHs could grow in mass through gas accretion and galaxy mergers confirmed that by the age of 900 million years, SMBHs with masses of $10^9$ $M_\odot$ and host galaxies with masses of $10^{12}$ $M_\odot$ (similar to the Milky Way) could form (Li et al. 2007).

### *5.2 Phase ②: Rapid Decline of AGNs (Factor (A) << Factor (B))*

Phase ②, the rapid decline of AGNs, is assumed to be the period during which AGNs formed in Phase ① lose their activity (corresponding to the period of rapid decline in Figure 13). The accretion disk heats up due to friction between gas particles, reaching extremely high temperatures of hundreds of thousands Kelvin, emitting strong X-rays and ultraviolet radiation from the central region of the disk. The outer regions of the disk cool down to temperatures of several thousand Kelvin, emitting visible light. Additionally, radio galaxies emit radio jets composed of electron-positron plasma flows. As the gas in the central region of the host galaxy is converted into electromagnetic energy and consumed by the SMBH, its mass increases, leading to the gradual disappearance of the gas. Over time, as the gas is depleted, the mass accretion rate (dM/dt) and luminosity ($L_Q$) decrease according to Equation 3, resulting in a decline in AGN activity.

Trend (2), "The number density of SFGs and AGNs reverses at z=1–3," corresponds to the increase in SFGs around z~2, as shown in Figure 15. Similarly in Figure 1, corresponds to the period around z~2 when the number of dim galaxies increased.

### *5.3 Phase ③: Gradual Decline of AGNs (Factor (A) < Factor (B))*

In Phase ③, the gradual decline of AGNs, the process of AGNs losing their activity and transforming into low-luminosity galaxies continues, but the decrease in AGNs becomes more moderate due to the merging of SFGs formed again in Phase ②, which transform into AGNs. The rate of decline in galaxy luminosity is expected to decrease (corresponding to the period of slower decline in Figure 13).

In the nearby universe, located only hundreds of millions of light-years away from Earth, many ultra-luminous infrared galaxies (ULIRGs) have been found, emitting infrared luminosity equivalent to $10^{12}$ $L_\odot$ (as bright as quasars) in the wavelength range of 1 μm–1000 μm. Many of these galaxies exhibit structures with tails (arms) extending from the main body of the galaxy. These tails are characteristic of galaxies that are merging or have already merged, leading to the conclusion that ULIRGs are merging galaxies (Taniguchi et al. 2012).

When spiral galaxies containing large amounts of gas merge, the gas within each galaxy collides violently. As the merger progresses, the gas falls further into the central region, triggering ultra-starbursts. The young massive stars formed during this process emit strong ultraviolet and visible light. The dust, heated by the absorbed energy, re-emits strong infrared radiation, making the galaxy appear as an abnormally bright infrared source. This state is interpreted as a ULIRG. On the other hand, young massive stars undergo supernova explosions, dying in a process known as supernova bursts. The galactic winds generated during these explosions blow away the gas and dust from the merging galaxies, revealing the bright central nucleus that emits X-rays, ultraviolet, and visible light. This phenomenon is explained as quasars and other AGNs (Caltech model: Sanders et al. 1988). Additionally, collisions between spiral galaxies and satellite galaxies lead to the formation of Seyfert galaxies (Taniguchi et al. 2012).

Using the time constant $\tau_x$ for SFGs to transform into AGNs, and applying Equation 4 along with $\tau_1$ = 900 million years and $\tau_2$ = 1.7 billion years from Figure 13, $\tau_x$ = -1.9 billion years was derived. The obtained value of 1.9 billion years is consistent with the order of magnitude of the galaxy merger period (approximately 3 billion years) reported by the NASA Hubble Mission Team in 2012.

### *5.4 Elucidations of the Problems for Dead Quasar and Downsizing*

The SMBHs that serve as the energy source for quasars do not die within the current age of the universe, approximately 13.8 billion years, however, quasars are rarely observed in the local universe. The absence of quasars can be understood as "SMBHs still exist within quasars, but due to the depletion of gas in the central regions of galaxies, accretion phenomena have ceased, and they are observed as ordinary galaxies (SFGs)."

As shown in Figure 1, the peak number density of bright (low absolute magnitude) galaxies is at high redshifts (early universe), making it appear as though larger, brighter galaxies formed earlier. This seems to contradict the idea of galaxy evolution through mergers. However, this can be explained by considering that "after quasars are formed, the decrease in gas accretion rates causes them to dim, leading to an increase in the number of dim AGNs while the number of bright AGNs decreases, each having its own peak." The fact that the number density of bright galaxies is lower is because brighter galaxies are converted into larger survey volumes. The similarity between Figure 1 and Phase ② (Rapid Decline of AGNs) in Figure 15 is also noteworthy.

## 5. SUMMARY

Using the BPT diagram with a fixed boundary method, by classifying active galactic nuclei (AGNs), which emit strong electromagnetic radiation from the central regions of galaxies, and star-forming galaxies (SFGs), which do not exhibit such activity, it was found that SFGs are distributed at low redshifts (z), while AGNs are distributed at high redshifts. The number density of SFGs and AGNs reverses between z=1 and z=3. Additionally, it was observed that the luminosity of galaxies decreases monotonically with the increase in cosmic age, indicating that galaxies were brighter when they were newly born and became dimmer over time.

Based on the above findings, a new model was proposed in which SFGs merge to form quasars in the early universe, followed by a decrease in gas near the center, leading to a weakening of activity and transformation into SFGs. In the nearby universe, SFGs merge again to form AGNs. This model successfully explains the "dead quasar problem" and "downsizing."

## ACKNOWLEDGEMENTS

I would like to express my deep gratitude to Professor Yoshiaki Taniguchi of the Open University of Japan for his continuous guidance and encouragement throughout this research. I also sincerely thank Associate Professor Takashi Murayama of Tohoku University for his careful advice and for introducing numerous valuous databases and papers.

I am deeply saddened by the passing of Professor Taniguchi on 9 January 2026 and respectfully offer my heartfelt condolences.